\documentclass[a4paper,11pt,aps,prx,amsmath,amssymb,amsfonts,superscriptaddress,notitlepage,reprint,longbibliography]{revtex4-2}

\usepackage{graphicx} 
\usepackage{hyperref}
\usepackage{lipsum}
\usepackage{color} 
\usepackage{lmodern} 
\usepackage{textcomp} 
\usepackage[T1]{fontenc} 

\DeclareMathOperator{\arccosh}{arccosh}
\newcommand*{\ket}[1]{|{#1}\rangle}

\def\be{\begin{equation}}
\def\ee{\end{equation}}
\def\bes{\begin{equation*}}
\def\ees{\end{equation*}}

\newcommand{\Qone}{$\mbox{Q}_1$\,} 
\newcommand{\Qtwo}{$\mbox{Q}_2$\,} 
\newcommand{\Pin}{P_{\rm in}}

\newcommand{\eq}{Eq.~}

\newcommand{\fig}{Fig.~}

\def\app{App.~}

\begin{document}

\title{Extensible quantum simulation architecture based on atom-photon bound states in an array of high-impedance resonators}

\author{Marco Scigliuzzo}
\email{marco.scigliuzzo.physics@gmail.com}
\affiliation{Department of Microtechnology and Nanoscience, Chalmers University of Technology, 412 96 Gothenburg, Sweden}
\author{Giuseppe Calajò}
\affiliation{ICFO-Institut de Ciencies Fotoniques, The Barcelona Institute of Science and Technology, 08860 Castelldefels (Barcelona), Spain}
\author{Francesco Ciccarello}
\affiliation{Universit\'a degli Studi di Palermo, Dipartimento di Fisica e Chimica, I-90123 Palermo, Italy}
\author{Daniel Perez Lozano}
\author{Andreas Bengtsson}
\affiliation{Department of Microtechnology and Nanoscience, Chalmers University of Technology, 412 96 Gothenburg, Sweden}
\author{Pasquale Scarlino}
\affiliation{Institute of Physics, Ecole Polytechnique Federale de Lausanne, 1015 Lausanne, Switzerland}
\author{Andreas Wallraff}
\affiliation{Department of Physics, ETH Z\"{u}rich, CH-8093 Z\"{u}rich, Switzerland}
\author{Darrick Chang}
\affiliation{ICFO-Institut de Ciencies Fotoniques, The Barcelona Institute of Science and Technology, 08860 Castelldefels (Barcelona), Spain}
\author{Per Delsing}
\affiliation{Department of Microtechnology and Nanoscience, Chalmers University of Technology, 412 96 Gothenburg, Sweden}
\author{Simone Gasparinetti}
\email{simoneg@chalmers.se}
\affiliation{Department of Microtechnology and Nanoscience, Chalmers University of Technology, 412 96 Gothenburg, Sweden}
\affiliation{Department of Physics, ETH Z\"{u}rich, CH-8093 Z\"{u}rich, Switzerland}

\date{\today}

\begin{abstract}
Engineering the electromagnetic environment of a quantum emitter gives rise to a plethora of exotic light-matter interactions. In particular, photonic lattices can seed long-lived atom-photon bound states inside photonic band gaps. 
Here we report on the concept and implementation of a novel microwave architecture consisting of an array of compact, high-impedance superconducting resonators forming a 1 GHz-wide pass band, in which we have embedded two frequency-tuneable artificial atoms.
We study the atom-field interaction and access previously unexplored coupling regimes, in both the single- and double-excitation subspace. In addition, we demonstrate coherent interactions between two atom-photon bound states, in both resonant and dispersive regimes, that are suitable for the implementation of SWAP and CZ two-qubit gates. The presented architecture holds promise for quantum simulation with tuneable-range interactions and photon transport experiments in nonlinear regime.
\end{abstract}

\maketitle

\section{Introduction}
Quantum emitters coupled to structured photonic environments constitute both an emerging paradigm of quantum optics \cite{roy_rev_mod,lodahl2017chiral} and a promising platform for quantum information processing \cite{Zheng2013,PaulischNJP16,pichler2017universal,borregaard2019quantum} and quantum simulation of many-body physics \cite{CarusottoRMP13,NohRPP16,Darrick_rev_mod,sheremet2021waveguide}.
One-dimensional photonic lattices modify the electromagnetic environment, leading to the appearance of finite bands and band gaps in the energy spectrum. A major phenomenon in these systems is the formation of atom-photon bound states in the photonic band gaps \cite{Byk,John1990a,John1994,Longo2010,QNO,Lombardo2014,calajoBS,Tao,Sanchez-Burillo2019a,Bello2019, VDS_PRL21}. These states originate from the dressing of the atom with a photonic cloud that remains exponentially localized in its vicinity, thus inhibiting a full atomic decay.
In addition, when multiple atoms are coupled to the same photonic lattice to form atom-photon bound states, their interaction is mediated by the overlap of their photonic wavefunctions. Since the photonic localization length can be controlled by varying either the frequency of the atom or the strength of its coupling to the lattice, this architecture supports tuneable-range interactions, opening opportunities for quantum simulation of exotic spin models \cite{Douglas2015,Tudela2015sub,marco,TaoNJP2018} and quantum computing architectures with connectivity beyond nearest neighbor \cite{naik_random_2017}.


	\begin{figure*}%
		\includegraphics[width=\linewidth]{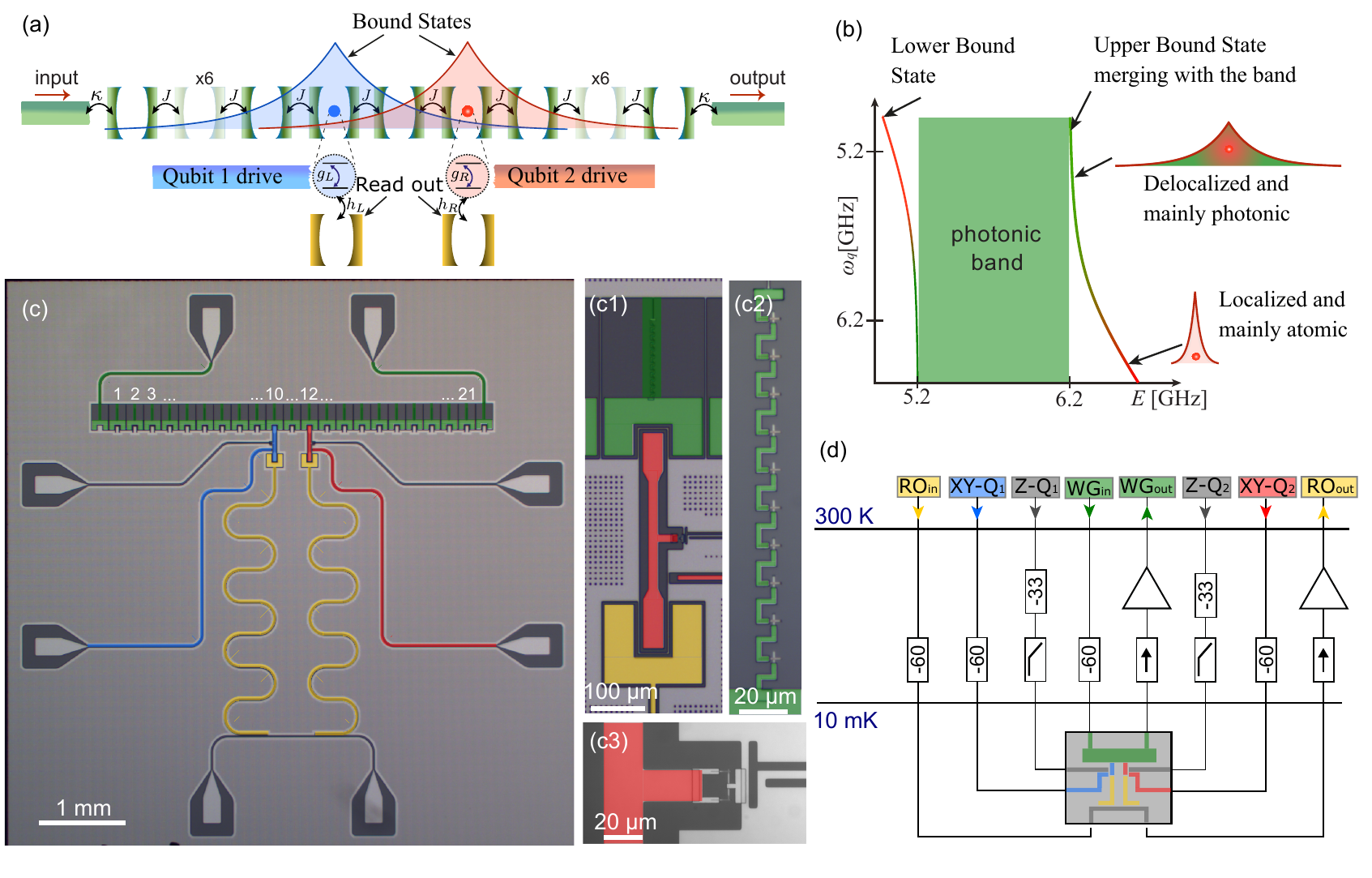}
		\caption{
			Two artificial atoms interacting with a coupled-cavity array. (a) Concept:  qubit 1, \Qone (blue) and qubit 2, \Qtwo (red),
			are locally coupled to an array of 21 cavities (green). Each atom can be independently 
			controlled through an individual driving line (same color as the atom) and its state measured via a dedicated read-out resonator (yellow). 
			Each qubit is dressed by a photon, giving rise to a corresponding atom-photon bound state (see shaded cloud localized around each atom).
			(b) Energy diagram of dressed states as a function of the bare qubit frequency $\omega_q$. For a fixed $\omega_q$, the bound state emerges as a discrete energy $E$ separated from the photonic band (green shaded region). The color scale of the bound state (from green to red) indicates whether the excitation is mostly photonic or atomic in nature. 
			(c) Micrograph of the sample and (d) simplified experimental setup: an array of 21 Josephson junction (JJ) resonators (green) is capacitively coupled at its edges to coplanar waveguides directly connected to the input (WG$_{\rm in}$) and output (WG$_{\rm out}$) of two cryostat lines. The transmon qubits Q1 and Q2 are coupled to resonator (cavity) 10 and 12, respectively, and
			driven through dedicated charge lines XY-\Qone and XY-\Qtwo, respectively. Each qubit state is detected through a
			two quarter wavelength coplanar resonator (yellow) inductively multiplexed on a transmission line measured through RO$_{\rm in}$ and RO$_{\rm out}$. The qubit frequencies are tuned by injecting a magnetic flux in the transmon SQUID loop with individual
			flux lines (Z-\Qone and Z-\Qtwo). Details of the JJ resonator and $Q_2$, the JJ array, and \Qtwo SQUID and flux line are shown in figure (c1), (c2) and (c3) respectively. 
		}
		\label{fig:concept}
	\end{figure*}




Atom-photon bound states have been observed in different systems, ranging from cold atoms coupled to photonic crystal waveguides \cite{KimblePNAS2016}, to optical lattices \cite{Krinner2018,Stewart2020}, to superconducting circuits \cite{houck,Sundaresan}.
A seminal experiment in superconducting circuits \cite{houck} relied on a microwave photonic crystal consisting of a coplanar waveguide with periodically modulated impedance. In a photonic crystal, the lattice periodicity must be of the same order as the wavelength of the radiation in the band gap. At microwave frequencies, this constraint results in a large footprint, which limits the number of unit cells that can be accommodated on a chip to a dozen and hinders the integration of additional measurement and control circuitry. 
A new avenue was opened by the introduction of superconducting metamaterials with a deep subwavelength lattice constant, consisting of a set of lumped-element resonators periodically loading a transmission line, or arranged in a linear chain to form a coupled-cavity array   \cite{painter1,ferreira2020collapse,paintertopo}. In these embodiments, the lattice footprint is drastically reduced (nearly two orders of magnitude) and the photonic cloud is more strongly confined. In addition, by staggering the hopping amplitudes between neighboring resonators, lattices with non-trivial topology have been realized, which host topologically protected bound states \cite{paintertopo}.

Here we introduce a novel circuit quantum electrodynamics (QED) implementation of photonic lattices coupled to quantum emitters, which employs arrays of high-impedance resonators and transmon qubits.
Compared to previous work, our implementation features a reduced footprint, an enhanced interaction strength, and a higher extensibility. These benefits are obtained by utilizing arrays of Josephson junctions as compact inductors in the resonator array.
We present spectroscopy and time-domain measurements of a proof-of-principle device comprising an array of 21 high impedance resonators and 2 transmon qubits with dedicated measurement and control circuitry. We characterize the mode structure of the array, the emergence of atom-photon bound states, their anharmonicity, the exchange and cross-Kerr interaction between two atom-photon bound states.
Our results demonstrate that the presented architecture is endowed with all essential building blocks to carry out quantum simulation and quantum information processing tasks, and to access nonlinear regimes of quantum optics. The foreseen possiblity to accommodate multiple emitters with a limited increase in physical footprint looks particularly promising for carrying out quantum simulations of large many-body spin Hamiltonians \cite{Douglas2015}.


\section{Results}\label{sec:Results}

\subsection{Sample and experimental setup}
We implement the structured photonic environment as a transmission line made out of 21 high-impedance microwave resonators forming a coupled-cavity array (\fig 1). Each resonator consists of an array of 10 Josephson junctions of total inductance $L_r=$8.87\,nH shunted by a capacitor $C_r=91.3$\,fF, resulting in a bare resonant frequency $\omega_r=5.593$\,GHz and a characteristic impedance $Z_r= 312\,\Omega$. Nearest-neighbor resonators are capacitively coupled to form a linear chain and each edge resonator is coupled to a 50 $\Omega$ coplanar waveguide.
The artificial atoms (\Qone and \Qtwo), are implemented as superconducting flux tunable transmons \cite{koch_charge-insensitive_2007} capacitively coupled to sites 10 and 12 of the array. Each transmon is additionally coupled to a charge line (XY control), a flux line (Z control), and a readout resonator.
The device is realized in aluminum on a silicon substrate with a standard lithographic process \cite{Burnett2019}. 
The sample is wire bonded to a copper sample holder thermally anchored to the mixing chamber stage of a dilution refrigerator at 10\,mK.
A summary of the relevant sample parameters and further experimental details are presented in \app \ref{app:Exp}.

\subsection{Transmission spectroscopy of the array}
\subsubsection{Bare coupled cavity array}
\begin{figure}%
\includegraphics[width=\linewidth]{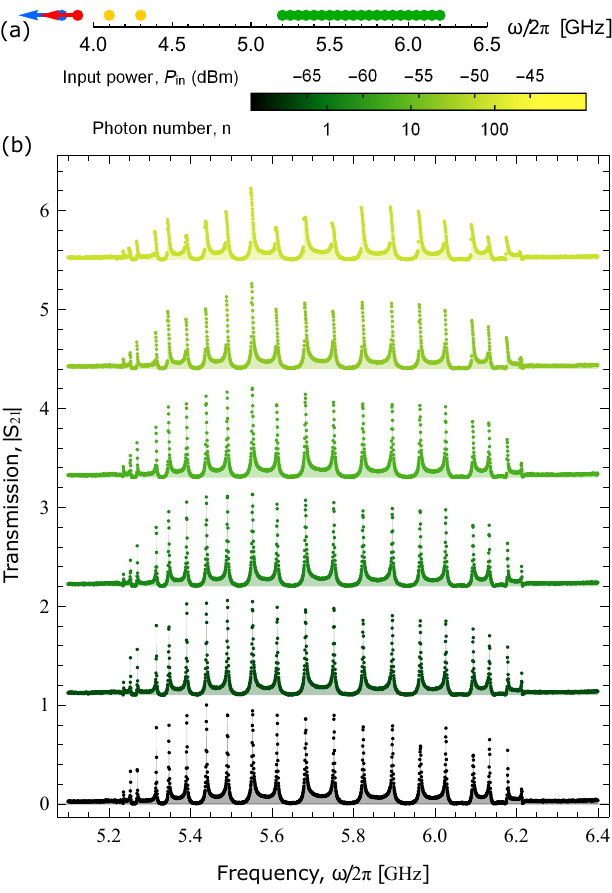}
\caption{Single coherent tone spectroscopy of the coupled cavity array. (a) Frequency landscape of the resonant modes of the systems: \Qone (blue), \Qtwo (red) (tuned away from the band in this measurement), readout resonators (yellow), and array modes (green). b) Transmitted amplitude, $|S_{21}|$ vs frequency, $\omega$, for different input powers, $\Pin$. The traces are vertically offset in steps of 1.1 and filled to their own baseline for clarity.
}
\label{fig:naked_array}
\end{figure}

To characterize the bare coupled cavity array, we tune the resonant frequencies of \Qone and \Qtwo far away from the transmission band and measure the transmission coefficient through the array
(\fig \ref{fig:naked_array}).
At low driving powers (average photon number in the array mode $n\approx 1$), a structure of $N=21$ modes forming a pass band emerges in the transmission spectrum, from the first mode at $\omega_1/2\pi=5.219$\,GHz to the last one at $\omega_{21}/2\pi=6.215$\,GHz. These modes present rather uniform spacing and linewidths in the center of the band, while they concentrate and become narrower in linewidth at the band edges. These features are captured by a tight binding model for the array with only three parameters: the single-cavity bare resonance frequency, $\omega_r$, the cavity-cavity nearest-neighbor coupling, $J$, and the coupling of the edge cavities to the input and output transmission lines, $\kappa$ (see \app \ref{app:theory}). In the limit $\kappa\ll J$, which applies here, the resonance frequencies of the modes are given by 
\begin{equation}
\label{eq:dispersion_main}
 \omega_k=\omega_r+2J\cos{\left(\frac{\pi k}{N+1}\right)}\;\;k=1,...,N\,,
\end{equation}
the linewidth of each being
\begin{equation}
  \kappa_k=\frac{2\kappa}{N+1}\sin^2{\left(\frac{k}{2}\right)}.
\end{equation}
\eq \eqref{eq:dispersion_main} predicts a pass band of width $4J$ centered around the bare cavity frequency $\omega_r$. 
From Eq.~\eqref{eq:dispersion_main},  $4J\approx\omega_{21}-\omega_1$ and we extract $J/2\pi=249$\,MHz. Comparing it to previous realizations of microwave coupled cavity arrays \cite{ferreira2020collapse,paintertopo}, we achieve a larger value for $J$ with a smaller coupling capacitance, thanks to the higher impedance of the resonator ($J=\frac12 C_J\omega_r^2 Z_r$). From electrostatic simulations we estimate the resonator impedance to be $Z_r=1/\omega_r C_r\approx 312\,\Omega$. This implies a sixfold gain on the capacitive coupling strength compared with 50\,$\Omega$ resonators. 

Compared to the predictions of this simple model, the measured traces
present deviations in the frequency distribution of the modes, that we attribute to a 3\% scatter in the resistance of the fabricated Josephson junctions \cite{junctionsStat}, resulting in an estimated  standard deviation of $\delta\omega_r/2\pi=$25\,MHz in the frequencies of the bare (uncoupled) resonators in the array.
Importantly, thanks to the small ratio $\delta\omega_r/J=1/10$, this frequency disorder does not significantly affect the properties of the atom-photon bound states, as calculated by numerical diagonalization of the system Hamiltonian for various realizations of the disorder, and directly verified in our experiments below.
In addition, we measure a nonzero transmission outside the photonic pass band, and an alternating transmission background in between resonant modes, which we ascribe to direct cross-talk between the input and output ports of our sample box.
For a detailed treatment of these experimental imperfections, see  \app \ref{App. Imperfections}.

At higher powers, corresponding to $n\sim 10^2$, each individual mode exhibits the typical phenomenology of a Kerr resonator, due to the nonlinearity inherited by the arrays of Josephson junctions \cite{andersen_ultrastrong_2017}.
In our design, we estimate the Kerr coefficient for a single mode to be $K/2\pi=100$\,kHz, much smaller than the mode linewidth. In fact, the frequency shift produced by more than 100 photons is still smaller than a linewidth as visible in \fig \ref{fig:naked_array}(b). 
Importantly, this value can be determined by design, both choosing the number of junctions in a single resonator, or the number of unit cells in the resonator array. The extraction of this parameter, and the relation between the Kerr coefficient of individual resonators and that of the array modes, are discussed in \app~\ref{app:Exp_photon}. In the remainder of this work, we will only consider the linear regime of the coupled cavity array.

\subsubsection{Dressed coupled cavity array and atom-photon bound states}
\begin{figure*}[htp]
\includegraphics[width=\linewidth]{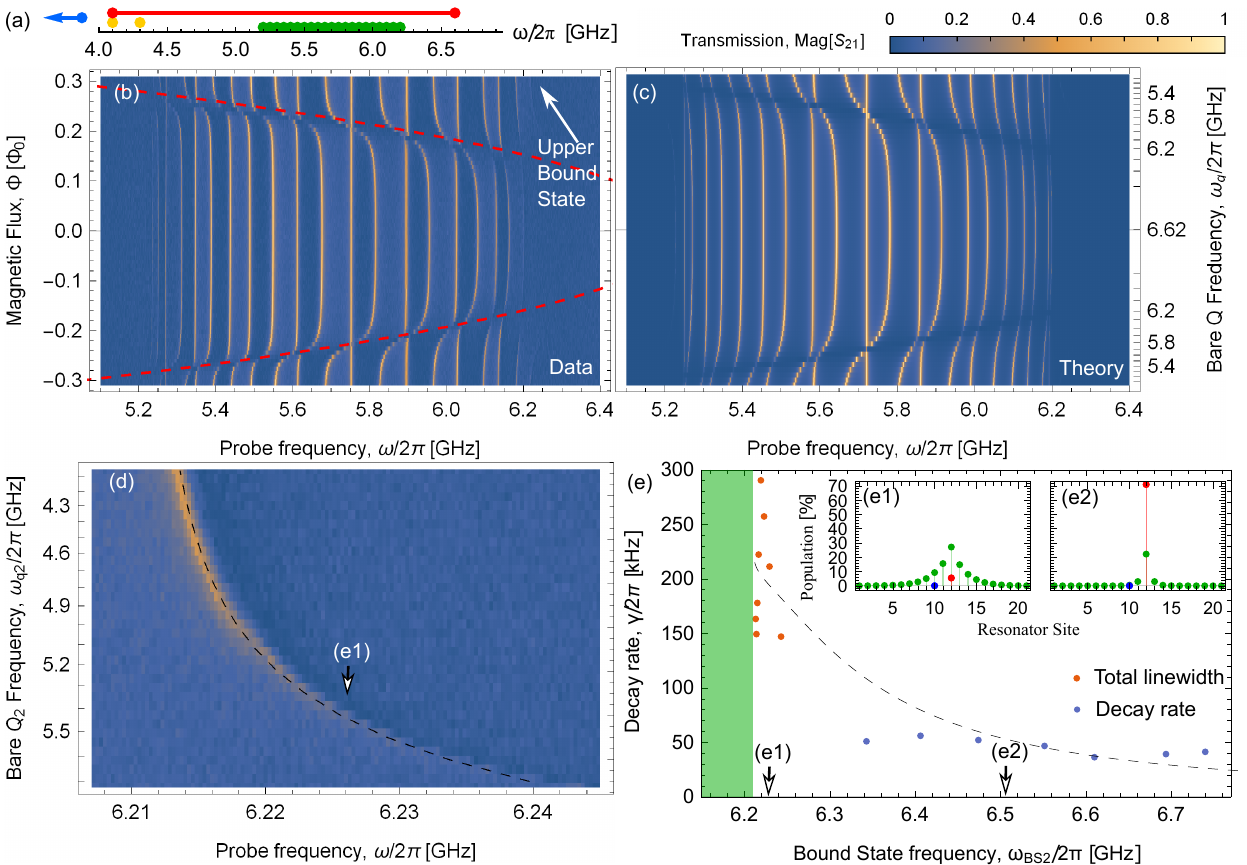}
\caption{Single coherent tone spectroscopy as a function of Q$_2$ frequency. (a) Frequency landscape: \Qone is tuned to its lowest frequency. The \Qtwo\, frequency is tuned from its largest frequency at 6.6\,GHz to 4.1\,GHz. b) Low-power coherent tone spectroscopy through the resonators array as a function of the \Qtwo\, frequency (dashed red line). (c) Calculated transmission spectrum via input-output theory using the single-excitation Hamiltonian \eq \eqref{eq:sys_hamiltonian}. (d) Detail of the measurement in (a) showing how the atom-photon bound state approaches the band and eventually becoming the last array modes. The dashed black line shows a data fit via \eq \eqref{eq:BSeig}. 
(e) Total decay rate of the bound state as function of its frequency. The red data are obtained from the transmission spectroscopy while the blue from the atom spectroscopy (see Sec.\ref{Sec.Sp_atom}). The dashed black line represents the theoretically expected decay rate.
Insets: calculated distribution of the excitation over the array (green dots) and between the two qubits (blue and red respectively) for two distinct  bound state  frequencies: $\omega_{BS2}=6.226$\,GHz (e1) and $\omega_{BS2}=6.510$\,GHz (e2).}
\label{fig:coherent_qubit}
\end{figure*}
We characterize the interaction of each qubit with the coupled cavity array by tuning its frequency across the pass band, while keeping the other qubit detuned [\fig \ref{fig:coherent_qubit}(a)]. Here we discuss the results for \Qtwo; the results for \Qone are comparable (see \app\ref{app:Q1}).

The low-power transmission coefficient of the array is affected by the presence of the qubit [\fig \ref{fig:coherent_qubit}(b)].
We observe a minimum in transmission within the band in correspondence with the bare qubit frequency (red dashed line in \fig \ref{fig:coherent_qubit}(b)). In fact, in the low excitation regime all the incoming field is coherently scattered back \cite{Fan}. This effect is particularly visible when the bare qubit frequency is resonant with one of the coupled-cavity array modes. In Sec. \ref{Sec.Sp_atom} we show an atom-cavity interaction strength is approximately $g_i/2\pi\sim 300$\,MHz, five times larger than the frequency spacing of the modes, implying a multimode interaction. In particular, the absence of an avoided crossing between a single mode and the artificial atom, and the monotonic dispersive shift of a mode,
indicate the interaction of the artificial atom with multiple modes.  Moreover each mode presents a definite standing-wave spatial profile, which sets its effective interaction strength with the artificial atom. 
For example, mode $m=10$, indicated by the white arrow in \fig \ref{fig:coherent_qubit}(b) and (c), is completely decoupled from the qubit due to a corresponding node on the site of the artificial atom.

Comparing our system with a previous realization of multimode strong coupling \cite{SundaresanPRX15}, the vanishing group velocity, $v_{\rm g}$, at the band edges produces a nontrivial density of photonic states, proportional to $1/v_{\rm g}$ \cite{manga_rao_single_2007}. As the atom frequency approaches the band, it hybridizes with band-edge photons having zero velocity \cite{houck, calajoBS}. This seeds an evanescent field, exponentially localized around the atom (shaded area around the atoms in \fig \ref{fig:concept}(a)). In \fig \ref{fig:coherent_qubit}(d) we observe this additional photon-like mode outside the passband, which asymptotically approaches the band edge when the bare qubit frequency is moved towards the center of the band. 



To model the transmission spectra, we introduce the Hamiltonian 
\begin{equation}
\label{eq:sys_hamiltonian}
\begin{split}
    H/\hbar= & \sum_{x=1}^{N}\omega_r a_x^\dagger  a_x
    +\sum_{x=1}^{N-1}J\left( a_x^\dagger  a_{x+1}+ a_{x+1}^\dagger  a_{x}\right)\\
    + & \sum_{i=1}^{2}\omega_{qi} b_i^\dagger  b_{i}+\frac{1}{2}\beta_i b_i^\dagger b_i^\dagger b_i  b_i+g_i\left( a^\dagger_{x_i} b_i+ b_i^\dagger  a_{x_i}\right), \\
    \end{split}
\end{equation}
where we introduce the photon annihilation operator $a_x$ for the x-th cavity, the transition frequencies $\omega_{qi}$ of qubits $Q_i$, their anharmonicities $\beta_i$, and finally their couplings with the $x_i$-th cavity, $g_i$.
We calculate the transmission coefficient in the limit of linear response from input-output theory
and find a qualitative agreement with our measurements [\fig \ref{fig:coherent_qubit}(c); see \app \ref{app:input-output} for details].  


When the frequency of the qubit is tuned towards the low frequency edge of the band, the bound state at higher frequency (upper bound state) completely loses its atomic nature, becoming the mode at the high frequency edge of the band. Conversely, the mode at the low frequency edge starts to get dressed with an atomic component (lower bound state, barely visible in \fig \ref{fig:coherent_qubit}(b)), as predicted by the Hamiltonian \eqref{eq:sys_hamiltonian} for a single qubit.

Their frequencies $\omega_{BSi}$ ($i=1,2$ for \Qone and \Qtwo, respectively), are given by the solutions of the equation 
 \begin{equation}\label{eq:BSeig}
\omega_{BSi}-\omega_{qi}=\frac{g^2}{(\omega_{BSi}-\omega_{r})\sqrt{1-\frac{4J^2}{(\omega_{BSi}-\omega_{r})^2}}},
\end{equation}
with $\omega_{BSi}-\omega_r>2J$ (see \app \ref{App.BS1atom1ex} for more details).
The upper bound state will be our main focus for the rest of the paper, and we will use its frequency as the independent variable to describe our measurements.
As pictorially shown in \fig \ref{fig:concept}(b), this state is highly localized and atom-like in nature for atomic bare frequency deep in the gap while it is more photon-like and delocalized for atomic frequency close to the pass-band. These features are directly exploited for exciting the bound state: the finite overlap of the photonic cloud with the resonators at the edges of the array allows for the detection of this state in transmission (see \fig \ref{fig:coherent_qubit}(b)-(d)).

 The decay rate of the bound state, extracted from the linewidth of the transmission spectroscopy as function of the bound state  frequency, is of the order of $\gamma/2\pi\approx300$\,kHz, close to the decay rate of the array modes [\fig \ref{fig:coherent_qubit}(e), orange dots], due to the large and delocalized photonic component of the bound state [\fig  \ref{fig:coherent_qubit}(e1)]. When the bound state is far from the band, we extract its decay rate from a measurement of its atomic component, via the readout resonators. In this case, the losses are much smaller [\fig \ref{fig:coherent_qubit}(e)] due to the mainly atomic component [\fig  \ref{fig:coherent_qubit}(e2)], and we measure a decay rate $\gamma/2\pi\approx 50$\,kHz.

\subsection{State preparation of a single atom-photon bound state}\label{Sec.Sp_atom}


\subsubsection{Single-excitation subspace}\label{Sec.1ex_sp}
\begin{figure}
\includegraphics[width=\linewidth]{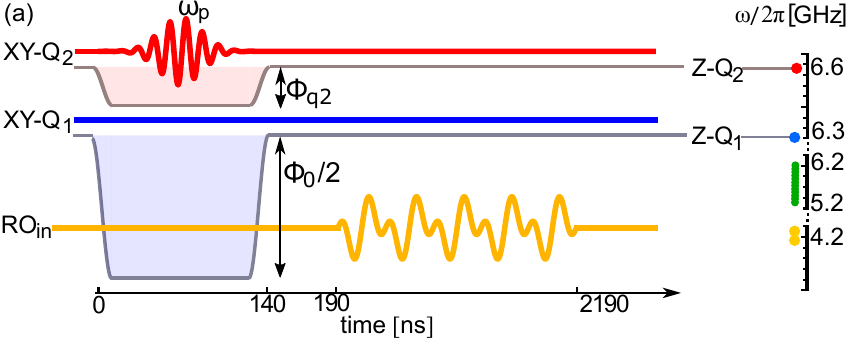}
\includegraphics[width=\linewidth]{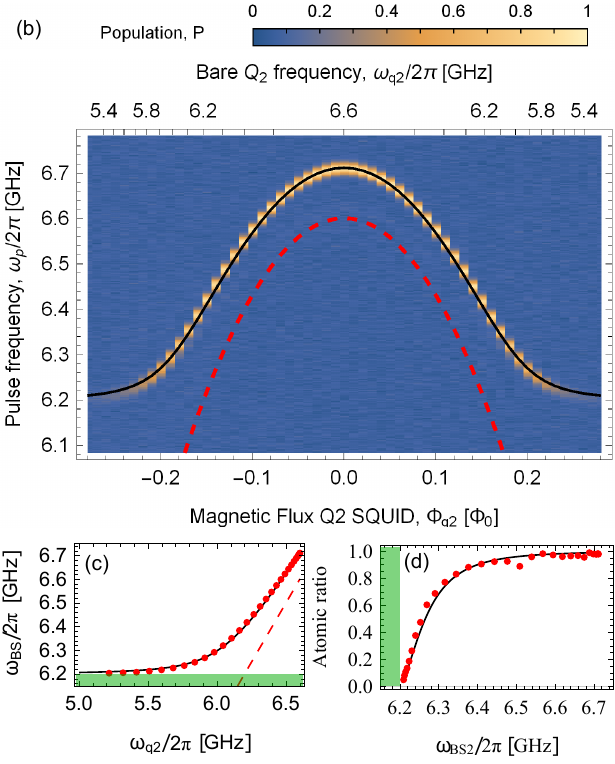}
\caption{Isolated bound state in single excitation subspace. (a) Pulse scheme. A Gaussian pulse (red line) with frequency $\omega_p$ drives \Qtwo, while \Qone is far detuned by flux pulse (blue shaded area). 
The readout pulse (in yellow), reads the population in both qubits. On the side,
\Qone and \Qtwo (blue and red dots, respectively), the photonic band (green dots) and the readout resonators (yellow dots) frequencies are depicted in relation to the flux pulses. (b) Population of \Qtwo as a function of flux pulse amplitude, $\Phi_{q2}$, and driving pulse frequency, $\omega_p$. The black line shows the fit of the bound state frequency given in \eq \eqref{eq:BSeig} as a function of the expected  bare \Qtwo frequency (red dashed line). (c) Bound state frequency extracted from panel (b). (d) Atomic fraction of the excitation. The black line shows the expected value (Eq. \eqref{theta}) using the parameters extracted from the fit in (b)-(c). 
}
\label{fig:BS_spec}
\end{figure}

In the limit in which the atom-photon bound state is localized, and thereby not accessible by transmission spectroscopy of the array, we excite it by sending microwave pulses to the qubit via its charge line, and detect it by performing dispersive qubit readout using the readout resonators [see \fig \ref{fig:concept}(c1) and pulse scheme in \fig\ref{fig:BS_spec}a].
A 90\,ns Gaussian pulse with frequency $\omega_p$ and calibrated amplitude to be $\pi$-pulse when \Qtwo is at its maximum frequency, $\omega_{q2,m}$, is sent to the XY-\Qtwo control line (red). At the same time, the frequency of \Qtwo, $\omega_{q2}$, is set with a 140\,ns square pulse (with 2\,ns rise and 2\,ns fall time) on the magnetic flux control, Z-\Qtwo with varying amplitude $\Phi_{q2}$ (grey line and red shaded area). The frequency of \Qone is dynamically tuned to its lowest value with a flux pulse of half flux quantum $\Phi_0/2$. After 50\,ns a 2\,$\mu$s square readout pulse sent to the multiplexed resonators, is used to read out the average qubit population, $P_2$.
The excitation spectrum of the bound state is obtained by measuring $P_2$ as a function of pulse frequency and magnetic flux applied to \Qtwo [\fig \ref{fig:BS_spec}(b)].
When the pulse on XY-\Qtwo is resonant with the BS transition frequency, the qubit is efficiently excited. The bound state frequency strongly differs from the bare qubit frequency [red dashed line in \fig \ref{fig:BS_spec}(b)], and asymptotically approaches the band edge for qubit frequencies in the band [\fig \ref{fig:BS_spec}(c)]. The bound-state frequency is also in this case well described by the continuum theory, as shown by the solid black line in \fig \ref{fig:BS_spec}(b), representing the best fit of Eq. \eqref{eq:BSeig} to the data. 
We notice that, keeping the pulse amplitude constant,
the bound state population decreases as its frequency approaches the band edge.
Compared to the drive rate $\Omega_{R,0}$ for an isolated qubit excited via its charge line by a pulse of given amplitude, the bound state is subject to a reduced drive rate $\Omega_R=\Omega_{R,0} \cos(\theta)$, where $\theta$ is the mixing angle between the atomic and photonic components. 
This relation allows us to extract the mixing angle from the measurements in \fig \ref{fig:BS_spec}(b), which we find to be in good agreement with our theoretical prediction [\fig \ref{fig:BS_spec}(d)].



\subsubsection{Double-excitation subspace}\label{Sec.twoex}

\begin{figure}%
\includegraphics[width=\linewidth]{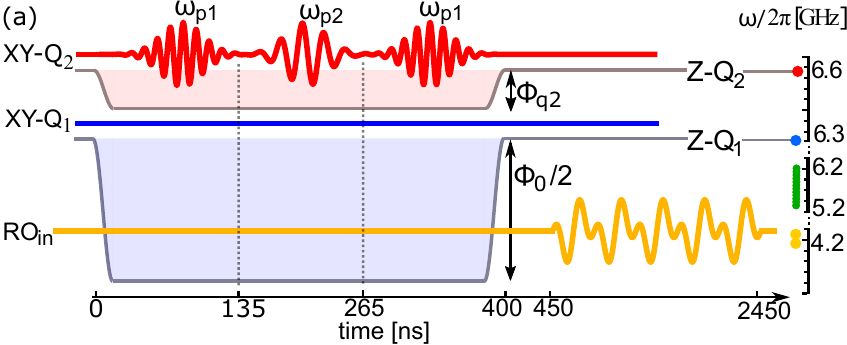}
\includegraphics[width=\linewidth]{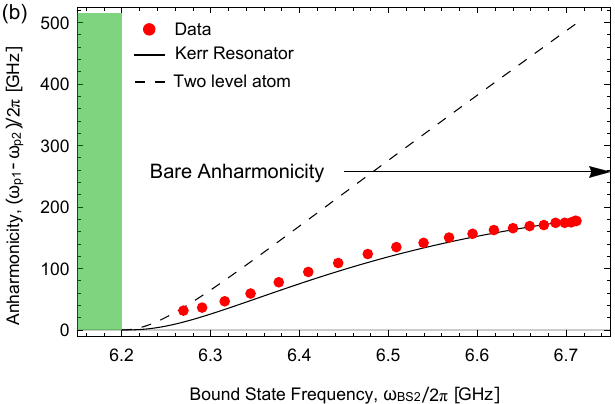}
\caption{Isolated bound state in the two-excitation subspace. (a) Pulse scheme.
A $\pi$-pulse (red line) excites \Qtwo, a second Gaussian pulse brings it to the third level when $\omega_{p2}$ is resonant with the ef transition. When $\omega_{q2}$ is not resonant, a $\pi$-pulse brings the qubit back to the ground state. (b) Absolute value of the dressed anharmonicity $\beta_{\rm dress}$. The arrow indicates the value of the bare anharmonicity. The black solid line shows the calculated $\beta_{\rm dress}$ with the parameters extracted from the fit in \fig \ref{fig:BS_spec}(b). Finally the black dashed line indicates the dressed anharmonicity of an ideal two-level atom.}
\label{fig:BS_Anharmonicity}
\end{figure}

The physics of the bound states explored so far was restricted to a single excitation.
When higher-excitation subspaces are considered, the nonlinear nature of the transmon starts to play a role and leads to deviations from the linear regime.
To explore these nonlinear features, we "climb up" to the two-excitation subspace using a sequence of two pulses [\fig \ref{fig:BS_Anharmonicity}(a)].
We first send a resonant $\pi$-pulse
to excite the long-lived, single-excitation bound state. We subsequently send another pulse of fixed amplitude and varying frequency $\omega_{p2}$ to search for the first-to-second-excited state transition of the bound state.
Finally, because our readout pulses are optimized so that the ground state is the most distinguishable one, we find it convenient to
add a third pulse identical to the first, so that if the second pulse does not affect the bound state, then the system is brought back to the ground state. Such a scheme allows us not only to perform a complete spectroscopy of the second excitation subspace but also, compared to previous approaches \cite{Sundaresan}, to efficiently prepare the two excitation bound state.


In order to quantify the deviation from the linear case we define the  dressed anharmonicity parameter as $\beta_{\rm dress}=\omega_{BS2}^{(2)}-2\omega_{BS2}^{(1)}$ where the superscripts $(1)$ and $(2)$ stand for the number of excitations. Experimentally, this quantity is determined as the difference in frequency between the second and first pulses when the second transition is resonantly excited.
The measured $\beta_{\rm dress}$ is always negative; its magnitude is maximum when the bound state frequency is the farthest from the band edge, and monotonically decreases towards zero as the bound state frequency approaches the band edge [\fig \ref{fig:BS_Anharmonicity}(b), dots].
To theoretically capture the measured nonlinearity, 
we describe the transmon as a nonlinear resonator with bare anharmonicity $\beta/2\pi=-257$\,MHz (see Eq.~\eqref{eq:sys_hamiltonian}).
This quantity is estimated by performing the same measurement as in \fig \ref{fig:BS_Anharmonicity}(a) for $\omega_{q2}/2\pi=3.329$\,GHz, well below the band and the read-out resonator frequency.
We use the parameters $J$, $\omega_r$ and $g_1$ extracted from the fit of the single excitation bound state spectroscopy, to diagonalize the Hamiltonian in the single- and double-excitation subspace (see \app \ref{App.2BS}) and to calculate the dressed anharmonicity, finding a good agreement with the measured data [\fig \ref{fig:BS_Anharmonicity}(b), solid line]. 

The observed anharmonicity for the bound state is intermediate between that of a fully linear emitter, for which no anharmonicity would be observed, and that of the most nonlinear emitter, a genuine two-level atom ($\beta \to -\infty$), plotted for comparison as a dashed line in \fig \ref{fig:BS_Anharmonicity}(b). 
A detailed numerical study, presented in App.~\ref{App.2BS}, indicates that double-photonic, double-atomic, and hybrid excitations are all present across the frequency range considered.
In addition, we find that the localization length of the doubly excited bound state is also renormalized according to the nonlinearity of the emitter, as originally discussed in \cite{calajoBS}.

\section{Bound states Interaction}

\subsection{Two atom-bound states level splitting }\label{Sec:splitting}

\begin{figure*}
\includegraphics[width=\linewidth]{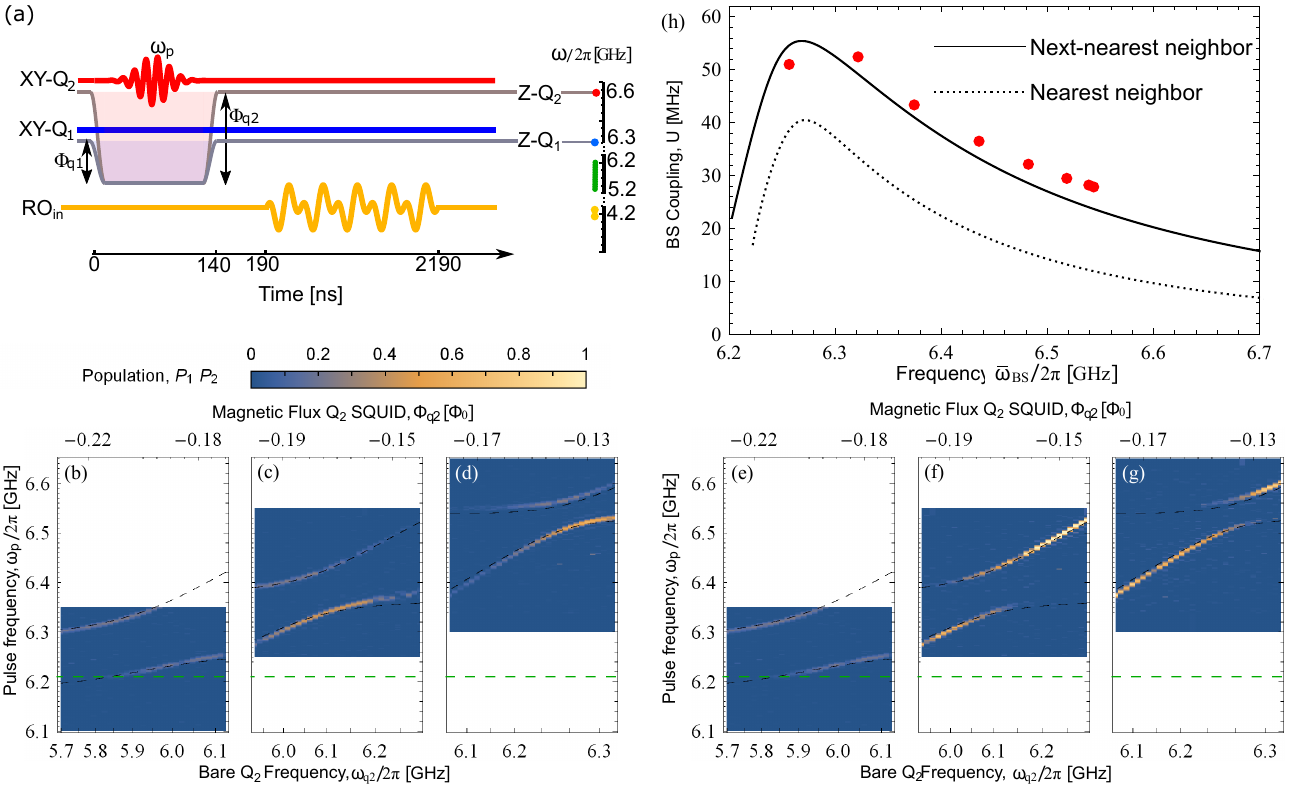}
\caption{Avoided crossing and interaction strength between two bound states. (a) Pulse scheme. A Gaussian pulse (red line) with frequency $\omega_p$ drives the two bound states tuned in resonance with two flux pulses (red and blue shaded areas). The populations of the qubits are read out (yellow line) at their highest frequencies. (b)-(g) Individual bound state population on \Qone [$P_1$ (b)-(d)]  and on \Qtwo [$P_2$ (e)-(g)] as a function of the bare frequency of \Qtwo for different frequencies of \Qone. The dashed black lines are the expected bound state frequencies calculated from Hamiltonian  \eqref{eq:sys_hamiltonian_full}. 
The green dashed line indicates the photonic band edge. 
(h) Bound states coupling $U$, as defined in the main text, extracted from the measurements reported in (b)-(g). The black dashed line represents the expected value calculated from the Hamiltonian \eq \eqref{eq:sys_hamiltonian} while 
the solid line is obtained by the extended model given in \eq \eqref{eq:sys_hamiltonian_full}, which takes into account the next-nearest neighbor interaction.
}
\label{fig:interaction}
\end{figure*}

To investigate the interaction between bound states, we bring their frequencies close to each other, send an excitation pulse to one of the qubits, and perform joint readout of the two qubits [\fig \ref{fig:interaction}(a)].
Sweeping the frequency of one qubit while keeping the other one fixed, we observe characteristic avoided crossings in the excitation spectrum, detected as peaks in the measured populations of both qubits as a function of the qubit and probe frequencies  [\fig \ref{fig:interaction}(b-g)].
Interestingly, the excitation pulse on \Qtwo, excites a population fraction in both bound states. In fact, when the two atom-photon bound states resonantly interact, they hybridize forming even $(+)$ and odd $(-)$ states (see App.\ref{App.2atom}). In \fig \ref{fig:interaction}(b-g) we report the measurement results for three different frequencies of \Qone: when it is tuned well inside the band $\omega_{q1}/2\pi\approx5.9$\,GHz, (b) and (e), at the band edge, $\omega_{q1}/2\pi\approx 6.2$\,GHz, (c) and (f), and finally at its largest bare frequency possible, $\omega_{q1}/2\pi=6.332$\,GHz (d) and (g).

Tuning the bound state frequencies closer to the band edge, results in a larger avoided crossing, corresponding to a larger interaction strength. It is important to notice that the two qubits do not present a direct coupling, in fact their interaction derives from the mutual overlap of their photonic clouds which becomes progressively more extended and populated closer to the band edge [see \fig \ref{fig:coherent_qubit}(e1)-(e2) where we calculated the photonic cloud distribution of the bound state depending on the distance from the band edge].

When the frequencies of the interacting bound states approach the band edge, we observe a vanishing population of the (-) bound state, when its frequency crosses the band edge [\fig \ref{fig:interaction}(b,e)]. This phenomenon, referred to as "melting" of one bound state into the modes of the band \cite{calajoBS}, only occurs in the presence of two interacting bound states, (+) and (-). The even state is characterized by a bonding behavior and its energy is raised compared to the one of the individual bound state, leading to an increased localization of its photonic cloud. The odd state instead presents an anti-bonding behavior, with its energy pushed towards the band edge making the state more extended.  This progressive delocalization leads to a disappearance of this bound state solution.

We define the bound states interaction strength $U=[\omega_{BS+}(\omega_q)-\omega_{BS-}(\omega_q)]/2$ as  half of the level splitting when the bare qubit frequencies are on resonance, $\omega_q=\omega_{q1}=\omega_{q2}$, where the $\pm$ stands for even and odd dressed bound states, respectively.
This quantity is displayed in \fig \ref{fig:interaction}(h) (red dots) as a function of the frequency at the midpoint of the level splitting, $\bar \omega_{BS}=\omega_{BS-}+U/2$, which gives the approximate frequency of the individual (single artificial atom) bound state.
We measure the bound states interaction strength ranging from 27 to 52 MHz in a frequency range of 250 MHz. This tunability is directly related to the variation of the overlap between the photonic clouds. Based on numerical simulation, we predict that the interaction strength could be tuned by two orders of magnitude through a straightforward optimization of the device parameters (see \app \ref{app:optimization}). 

\fig \ref{fig:interaction}(h) also shows that the coupling does not present a monotonic behavior. In fact, it increases for bound state approaching the band edge, but it exhibits a maximum close to the melting condition.
This behavior relies on our definition of the BSs coupling and on the melting condition. Indeed, when the odd bound state merges with the continuum in the band, the splitting is measured between the even bound state and the band edge. 

Separately, we find that the expected interaction strength calculated with the eigenvalue equation derived by the system Hamiltonian \eq \eqref{eq:sys_hamiltonian} (see  \app \ref{App.2atom} for more details) systematically underestimates the measured coupling strength [\fig \ref{fig:interaction}(h), dotted line].
As shown in \app \ref{app:parassitic}, parasitic capacitances are responsible for a non-negligible coupling between each qubit and their next-nearest array sites.
In particular, the bound-state interaction is affected by a direct coupling between the qubits and the cavity in-between them ($x=11$). After including these additional couplings in the Hamiltonian,
we obtain a much better agreement with the data [\fig \ref{fig:interaction}(h), solid line]. This behavior cannot be explained, in contrast, by adding just a direct qubit-qubit coupling to the model.

\subsection{Time resolved excitation exchange}
\begin{figure}%
\includegraphics[width=\linewidth]{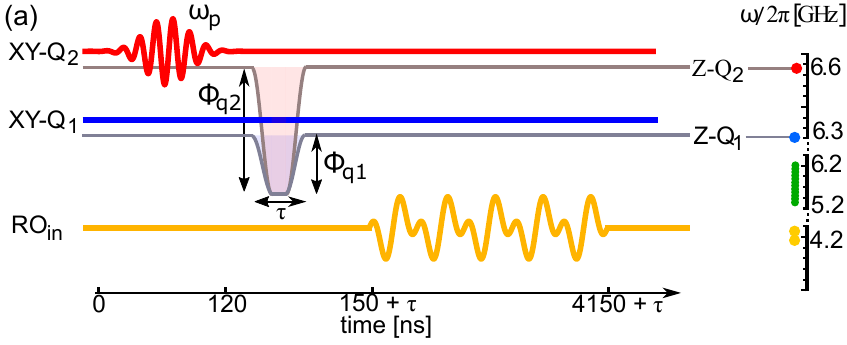}
\includegraphics[width=\linewidth]{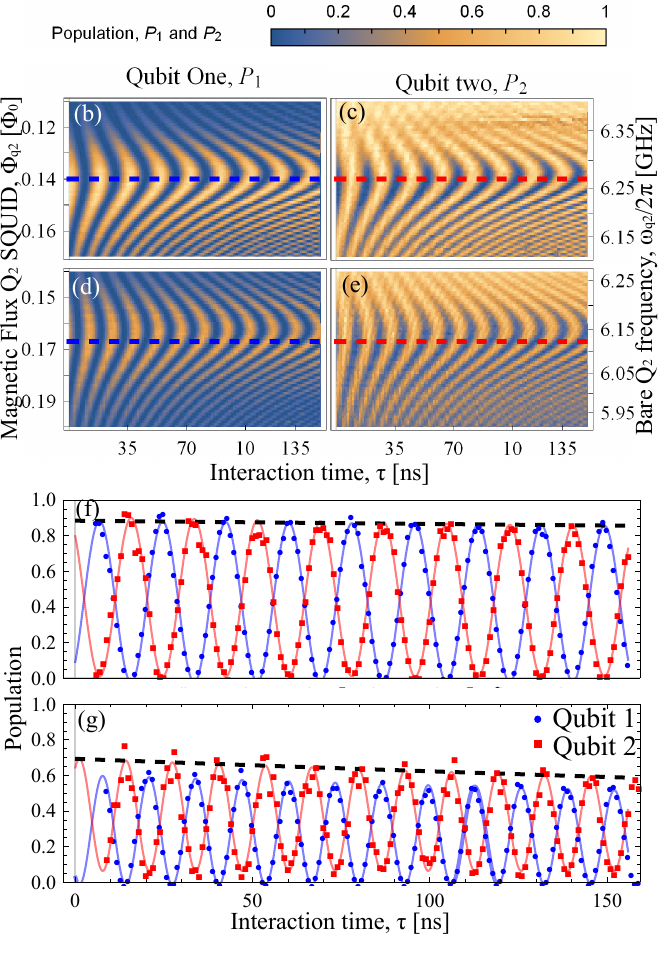}
\caption{Time resolved energy swap. (a) Pulse sequence. A $\pi$-pulse (red line) excites \Qtwo and after 30\,ns the two bound state tuned in resonance  with two flux pulses (blue and red shaded area) for a variable duration $\tau$. The qubits are tuned back to their unbiased frequencies and their populations are read out (yellow line). 
(b)-(e)Chevron pattern. Measured bound state population of \Qone (left column) and \Qtwo (right column) as a function of the interaction time and of the bare frequency of \Qtwo The two rows correspond to two different values of $\omega_{q1}$, at its largest frequency, (b)-(c), and 6.02\,GHz, (d)-(e), respectively. 
(f)-(g) Horizontal line cuts in panels (b)-(e) (blue and red dashed line) displaying the population of \Qone (blue) and \Qtwo (red) as function of the interaction time. 
In both plots the relaxation time is fitted to an exponential decay (black dashed line).}
\label{fig:time_interaction}
\end{figure}


After establishing a static interaction between the bound states, we now exploit it in a dynamic setting to realize an excitation swap between two bound states [\fig \ref{fig:time_interaction}(a)].
When both qubits are at their largest frequencies, we excite the bound state of \Qtwo with a $\pi$-pulse and we tune it in resonance with the bound state of \Qone, by applying a flux pulse on each individual flux line. After an interaction time $\tau$, we bring the two qubits back to their initial frequencies and we read-out the populations $P_1$ and $P_2$, in \Qone and \Qtwo, respectively. 

\fig \ref{fig:time_interaction}(b) and (c) show $P_1$ and $P_2$ as a function of the bound states interaction time, $\tau$, and \Qtwo frequency $\omega_{q2}$, while the \Qone bound state frequency is kept at $\omega_{q1}=\omega_{q1,m}$ in (b)-(c), and $\omega_{q1}/2\pi=6.15$\,GHz in (d)-(g). These are the same parameters chosen for measuring the avoided crossing in \fig \ref{fig:interaction}(d) and (g). 
The population of the two qubits as a function of the interaction time, restricted to resonant bound states, is presented in \fig \ref{fig:time_interaction}(f) and (g). These data, corresponding to a slice of the chevron pattern (dashed blue and red line in \fig \ref{fig:time_interaction} (b)-(e)), can be fitted to a dumped sinusoidal function (solid red and blue line), from which we extract a complete swap time of 18\,ns. This value is in good agreement with the measured interaction strength. 

The excitation swap between the two bound states is realized by non adiabatically tuning the two bare bound states in resonance. The adiabatic threshold for this process is related to the interaction strength \cite{solinas_decoherence_2010}, with a relative time scale  $\sim 1/U\sim 5$\,ns. The flux pulse we implement has a rise (and fall) time  of $t_{\rm raise}=1$\,ns, thus we fulfill this constraint. Nevertheless the two bound states interact during the rise time, and we take this into account with a time shift of the measured data in \fig \ref{fig:time_interaction}(f). 

The Chevron pattern of the population $P_1$ and $P_2$ measured for the bare frequency $\omega_{q1}/2\pi\approx6.02$\,GHz shown in \fig \ref{fig:time_interaction}(d)-(e) highlights a faster excitation swap of 13\,ns, as expected from the larger interaction strength closer to the band edge. Focusing only on the resonant case (see \fig \ref{fig:time_interaction}(g)) we notice that after the pulse protocol we observe a total population of only $P_1+P_2\sim 0.7$. The population loss originates from the weakly adiabatic regime in which we operate, i.e. $t_{\rm raise}\gtrsim 1/\Delta_{\bar \omega_{BS}}\sim 0.5\,ns$, where $\Delta_{\bar \omega_{BS}}$ is the gap between the average individual bound state frequency $\bar \omega_{BS}$ and the band edge (see Sec. \ref{Sec:splitting}).

When the bare qubit frequency is tuned in the band faster than the latter adiabaticity threshold, the bare qubit state is not an eigenstate of the system. In this case, part of the atomic population redistributes among the resonators according to the projection of the bare qubit state on the bound state. In this process, a fraction of the atomic population is converted into itinerant photons and released into the waveguide. In particular, the released population is at most $P_{\rm released}\simeq \sin^2(\theta)$ with  $\theta$ being the mixing angle $\theta$ defined in Sec.\ref{Sec.Sp_atom}. In fact, where the bound states are mainly qubit-like, the total population $P_1+P_2\sim 0.95$ (see  \fig\ref{fig:time_interaction}(f)), while a larger photonic dressing preserves a smaller fraction of population, $P_1+P_2\sim 0.7$, as shown in \fig \ref{fig:time_interaction}(g). This last value could be improved by slowing down the protocol but still keeping  $t_{\rm raise}\lesssim 1/U$ in order to induce the excitation swap in the first place.

The lifetime of the interaction is determined by the lifetime of the average individual bound state (dashed black lines in \fig \ref{fig:time_interaction}(f)-(g)), which we extract from the fit to be 1.2\,$\mu$s and 950\,ns, respectively.

\subsection{Two atom bound states ZZ interaction}


When the two atom photon bound states are close in frequency to the band edge, but detuned from each other, exciting one of them results in a shift in the transition frequency of the other. This ZZ (or cross-Kerr) interaction originates from dispersive and resonant interactions between energy levels in the double-excitation manifold. The interaction mechanism relies on the overlap of a photonic cloud with finite localization, as it happens in the single excitation subspace.

\begin{figure}%
\includegraphics[width=\linewidth]{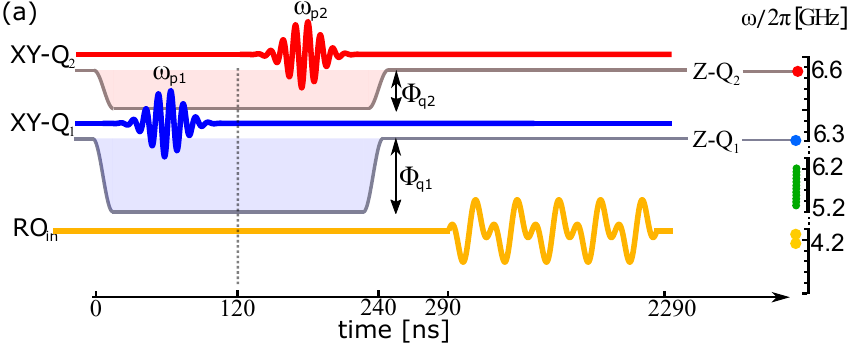}
\includegraphics[width=\linewidth]{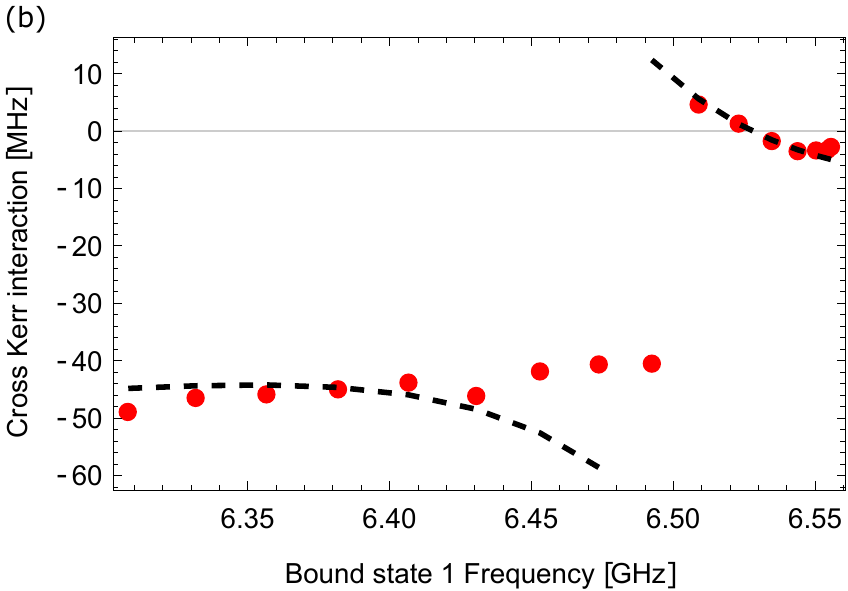}
\includegraphics[width=\linewidth]{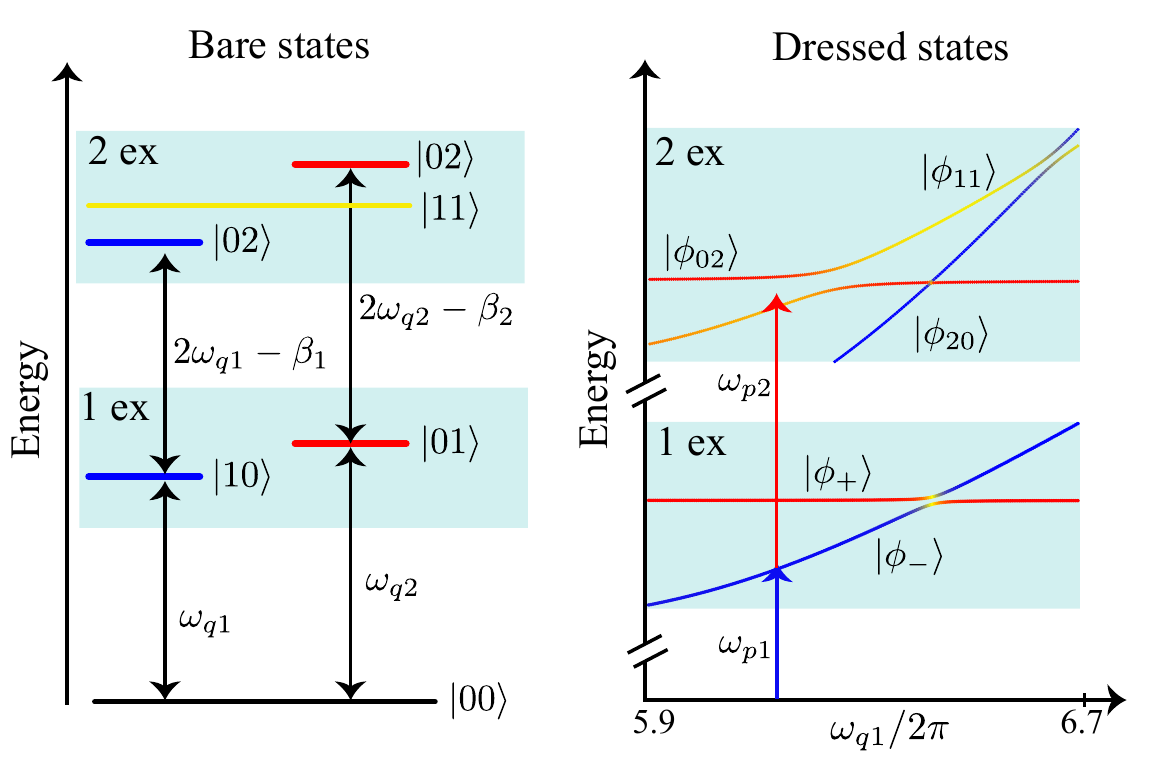}
\caption {Bound state interaction in two excitation subspace. (a) Pulse sequence. After tuning the bound states frequency with flux pulses (red and blue shaded area) they are excited (red and blue lines) and their population readout. (b) Cross Kerr interaction and avoided crossing between the states $|02\rangle$ and $|11\rangle$ for the frequency of the \Qtwo bound state $\omega_{\rm BS/2}/2\pi=6.653$\,GHz. The dashed line represent the expected value calculated with the Hamiltonian \eq \eqref{eq:sys_hamiltonian_full}. (c) Energy levels scheme for the bare  (left) and dressed (right) states. \label{fig:cross_kerr}
}
\end{figure}

We investigate the ZZ interaction with the pulse scheme displayed in \fig \ref{fig:cross_kerr}(a).
After setting the bare qubit frequencies with a flux pulse on each flux line, we excite \Qone bound state with a $\pi$-pulse. 
We apply a second pulse to \Qtwo with frequency
$\omega_{p2}$, and finally read out the state of the qubits. When the pulse on \Qtwo is such that $\omega_{p2}=\omega_{BS2}^{(2)}$, the bound state of \Qtwo is excited. 
We repeat the same pulse sequence by tuning the bare frequency of \Qone, mapping the transition frequencies $\omega_{BS2}^{(2)}$ as the function of $\omega_{q1}$.

\fig\ref{fig:cross_kerr}(b) shows the ZZ interaction between two excited bound states for the bound state of \Qtwo tuned at the constant frequency $\omega_{\rm BS/2}/2\pi=6.653$\,GHz, while the frequency of the \Qone bound state is tuned from band edge toward its largest frequency 6.55\,GHz. Following the level $|\phi_{11}\rangle$ on the right panel of \fig \ref{fig:cross_kerr}(c), we notice that this level crosses $|\phi_{02}\rangle$ relative to the two-excitation bound state of \Qtwo. In \fig\ref{fig:cross_kerr}(b), at lower frequencies,  we measure a residual ZZ interaction up to 49\,MHz. The dashed line shows the calculated ZZ interaction numerical evaluated with the complete Hamiltonian \eq \eqref{eq:sys_hamiltonian_full}. 

Note that in \cite{Sundaresan} the avoided crossing between $|\phi_{02}\rangle$ and $|\phi_{11}\rangle$ was infer using high power coherent spectroscopy. With this approach, second order transitions may induce Stark shift in the measured frequencies, while the measurement of the ZZ interaction with a pulse scheme is directly relevant for gate implementation.

\fig \ref{fig:cross_kerr}(c) shows the energy structure of bare qubits (left panel) and the same energy levels for bound states as function of the bare frequency of \Qone. The color scale highlights the population of the two atoms going from blue, for \Qone, to red for \Qtwo, where the yellow color instead stands for excitation equally distributed on the two. 
\fig \ref{fig:cross_kerr}(c) is a simplified version of the complete energy structure of our system. In fact, the three photonic bands that originate from the coupled cavity array, are not included in the figure (see \app \ref{App.2BS} for more details). 


\section{Discussion and Conclusion}

In summary, we introduced a new implementation of a finite band waveguide made out of an array of high impedance resonators to which we coupled two artificial atoms. We measured the transmission spectrum of the coupled resonators array, highlighting the underlying multimode interactions with the artificial atom, and observed the formation of the atom-photon bound states in the band gap. We demonstrated full control in accessing and preparing the atom-photon bound states in both the single and double-excitation subspace. We characterized the resonant interaction between two bound states in the static and dynamic regime, measuring an effective coupling strength up to 52\,MHz, and an excitation swap time down to $13\,$ns. Finally, we investigated the cross-Kerr (ZZ) interaction between two detuned bound states, reaching a value up to 49\,MHz. As shown in simulations, we expect a straightforward optimization of the device parameters to significantly improve our ability to control as well as suppress these interactions.

Comparing to previous implementations of superconducting qubits coupled to gapped waveguides, our approach based on high impedance resonators makes it possible to reach atom-cavity coupling strengths of a few hundred MHz, while maintaining the resonators footprint comparable to the one of the artificial atoms. Small footprint and strong interactions translate in a higher extensibility of our platform, with the foreseen possibility of adding more qubits as well as anchoring points to move towards two-dimensional lattices\cite{kollar_hyperbolic_2019}.

The tuneable-range interactions between atom-photon bound states available in this platform find application well to the quantum simulation of spin models \cite{Douglas2015}. At the same time, the possibility to implement fast and high-contrast SWAP and CZ gates using the array as a quantum bus could be further investigated in the context of quantum computing.
From the perspective of quantum optics, this platform is amenable to studies of 
correlated nonlinear photon transport \cite{Longo2010,LongoPRA11,HafeziPRA12} and quantum nonlinear optics protocols\cite{QNO}. By varying the coupling strength between neighboring sites, it is possible to engineer the band structure of the array and to endow it with nontrivial topological properties \cite{paintertopo}. The intrinsic nonlinearity of the array, whose strength can be adjusted by design, could be utilized to implement recent theory proposals describing exotic light-matter interaction effects \cite{wang2020supercorrelated,CarusottoPRL09,ChangNatPhys08}.
Finally, from the viewpoint of materials, replacing Josephson junction arrays with high-kinetic inductance superinductors \cite{samkharadze2016,niepce2019a} may lead to a more robust fabrication process and reduced disorder in the array.

\section*{Acknowledgment}\label{app:acknowledgement}
The authors are grateful to Peter Rabl, Angelo Carollo, and Christopher Warren for useful discussions.
MS, PD and SG wish to express their gratitude to Lars J\"{o}nsson for making the sample holder. They acknowledge financial support from the Swedish Research council and the Knut and Alice Wallenberg Foundation. 
DEC acknowledges support from the European Union’s Horizon 2020 research and innovation programme, under European Research Council grant agreement No 101002107 (NEWSPIN); the Government of Spain (Europa Excelencia program EUR2020-112155, Severo Ochoa program CEX2019-000910-S, and MICINN Plan Nacional Grant PGC2018-096844-B-I00); Generalitat de Catalunya through the CERCA program, Fundació Privada Cellex, and Fundació Mir-Puig.
The device was fabricated at Myfab Chalmers.

\appendix

\section{Experimental details}\label{app:Exp}

\subsection{Experimental Setup}
\begin{figure}%
\includegraphics[width=\linewidth]{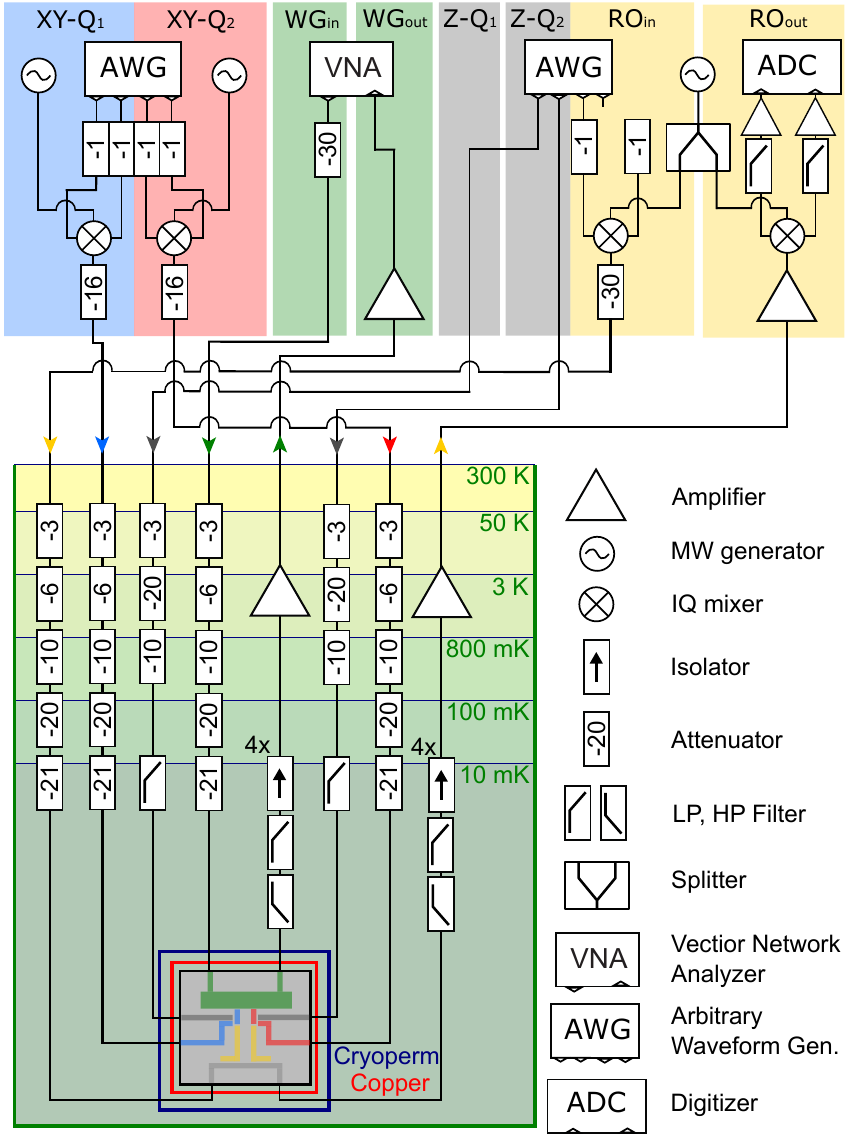}
\caption{Complete wiring diagram and room temperature setup. The coherent tone to the waveguide is generated and measured with a Vector Network Analyzer (VNA). \Qone and \Qtwo driving (blue and red respectively) are made by analog up-conversion of the pulsed generated by an arbitrary waveform generator (AWG). The flux lines are controlled by two AWG channels. Finally the qubits read-out (yellow) is controlled by thr up- and down-conversion of a pulsed generated by an AWG, and logged by an analog to digital converter (ADC).  
}
\label{fig:full_wiring}
\end{figure}

The complete experimental setup is shown in \fig \ref{fig:full_wiring}. The sample is wire-bonded in a nonmagnetic oxygen-free copper sample box (see \app \ref{app:crosstalk}), mounted to the mixing chamber stage of a dilution refrigerator operating at 10\,mK, and shielded by an additional copper and $\mu$-metal cans. Two additional shields (in order, copper and $\mu$-metal) are placed around the sample in a light-tight fashion. The signal to the \Qone and \Qtwo charge lines (XY-\Qone and XY-\Qtwo respectively), to the waveguide (WG$_{\rm in}$) and to read-out resonators (RO$_{\rm in}$) is delivered by an highly attenuated  coaxial lines (nominal total attenuation of -60\,dB). The flux lines (Z-\Qone and Z-\Qtwo) do not have any attenuation at the last two stages (total nominal attenuation -33\,dB) but they are equipped with a 4\,GHz low pass filter for reducing flux noise from higher stages. The transmitted signals (through the couple cavity array and through the readout feedline) is amplified with high electron mobility transistor (HEMT) amplifier at 3\,K, further amplified at room temperature. 

The microwave control for \Qone (blue area) and \Qtwo (red area), is achieved up-converting the inphase (I) and the quadrature (Q) components of a low frequency pulse generated by an arbitrary waveform generator (AWG), while the flux control is obtained injecting current generated directly by two AWG channels. The qubits readout pulse is up and down-converted with the same local oscillator (LO) (yellow area). Finally, the waveguide coherent spectroscopy is set up with a vector network analyzer (VNA) (green area).

\subsection{State preparation of \Qone}
\label{app:Q1}

\begin{figure}
\includegraphics[width=\linewidth]{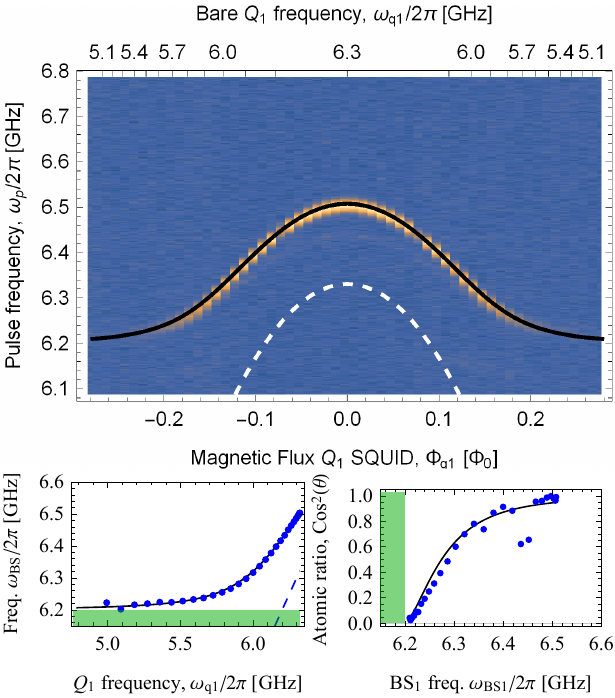}
\caption{State preparation of the isolated \Qone.  (a) Population of \Qone as a function of flux pulse amplitude, $\Phi_{q1}$, and driving pulse frequency, $\omega_p$. The black line shows the fit of the bound state energy given in \eq \eqref{eq:BSeig} to the measured data. The white dashed line shows the bare \Qone frequency calculated with the parameters extracted from the fit. (b) Bound state frequency extracted from panel (a) as function of the bare \Qone frequency. (c) Atomic fraction of the excitation, extracted from the relative population measured in the Qubit. The black line shows the expected value of the mixing angles as predicted from the ideal case theory given in Eq. \eqref{theta} using the parameters extracted from the fit in (a)-(b).}
\label{fig:BS_spec_q1}
\end{figure}

Using the same pulse scheme as the one shown in \fig \ref{fig:BS_spec_q1}(a) applied on \Qone we measure its population as a function of the bare qubit frequency (dashed white line) and driving pulse frequency, $\omega_p$. The result,  reported in \fig \ref{fig:BS_spec_q1}(a), shows \Qone largest frequency, $\omega_{q1,m}=6.332$\,GHz, smaller than \Qtwo, but compatible with the fabrication yields. The fit of the bound state frequency given by \eq \eqref{eq:BSeig} is in good agreement with the data (solid black line). The plot of the extracted $\omega_{BS,1}$ as a function of the bare qubit frequency is shown in \fig \ref{fig:BS_spec_q1}(b), and the atomic population fraction is reported in \fig \ref{fig:BS_spec_q1}(c), where the solid line is the calculated value with the parameters extracted from the fit.

\subsection{Photon number estimation}
\label{app:Exp_photon}
We calibrate the photon number in the array modes using the mode Kerr nonlinearity inherited by the Josephson junctions array resonators. The self-Kerr coefficient of the bare resonator scales with the inverse square of the number of junctions,  $K_r=e^2/(2\cdot 10^2\cdot C_r)=2\pi\times 2.1$\,MHz. Moreover the nonlinearity is further diluted in each mode as with an effective Kerr  $K= K_r/N=2 \pi\times100$\,kHz; in fact two excitations in one mode, in the momentum space, are spatially distributed to all the resonator of the array  \cite{weisl_kerr_2015}. We can experimentally confirm this estimation investigating the power response of the array modes.

\begin{figure}%
\includegraphics[width=\linewidth]{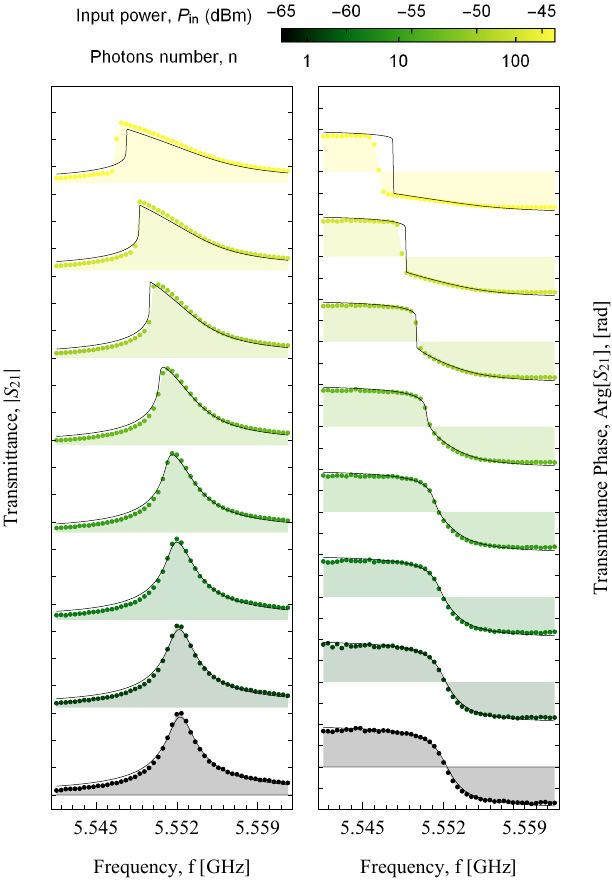}
\caption{Single coherent tone spectroscopy of a single mode. (a) Magnitude and (b) phase of the transmission of the mode of the coupled-cavity array at 5.552\,GHz as a function of power. The self-Kerr from the JJ array resonator is inherited by the modes. The coherent tone addresses only one cavity mode, so the global fit of a Kerr nonlinear resonator (continuous line) is used to extract the self-Kerr, $K$, and the photon number, $n$ of the array mode. The shade areas indicate the deviation zero transmission (deviation from zero phase). }
\label{fig:photon_calib}
\end{figure}

\fig \ref{fig:photon_calib} shows the phase and magnitude of a coherent microwave tone in the range of mode number 8 at 5.552\,GHz. Adapting the treatment in \cite{eichler_controlling_2014} to a transmission configuration, and assuming a total input attenuation at the sample equal to 70\,dB, we can globally fit the data to 
\begin{equation}
    S_{21}= \frac{\kappa}{\kappa_{\rm tot}}\frac{1}{1/2+ i\delta +i \xi \tilde n},
\end{equation}
where $\kappa$ is the external coupling rate, $\kappa_{\rm tot}$ is the total loss rate, $\delta=(\omega-\omega_0)/\kappa_{\rm tot}$ is the relative detuning, $\xi=(P_{\rm in}/\hbar\omega)\kappa K /\kappa_{\rm tot}^3$ is the (adimensional) input power, $K$ is the mode Kerr and finally $\tilde n$ is given by the solution of the algebraic equation
\begin{equation}
    1=(\delta^2+1/4)\tilde n-2\delta\xi\tilde n^2+\xi^3\tilde n^3.
\end{equation}
The global fit (solid black lines) reproduces the data well.

\subsection{System parameters}
The complete list of the system parameters measured in our sample is shown in Table \ref{tbl:parameters}. 
\begin{table}
\caption{Complete list of the system parameters.\label{tbl:parameters}}
 \begin{tabular}{l l} 
 Parameter & Value \\  
 \hline\hline
 \multicolumn{2}{c}{Resonators}\\
 \hline
 Resonator capacitance, $C_r$ & 91.3\,fF   \\
 \hline
  Resonator inductance, $L_r$ & 8.87\,nH   \\
 \hline
Resonator frequency, $\omega_r/ 2\pi$ & 5.593\,GHz   \\
  \hline
 Resonator impedance, $Z_r$ & 312\,$\Omega$ \\
  \hline
Non-radiative decay  $\gamma_r/2\pi$ & 300\,kHz \\
 \hline
Resonator self Kerr, $K_r/2\pi$ & 2.1\,MHz   \\
 \hline
 \multicolumn{2}{c}{Coupled Resonators Array}\\
 \hline
 Pass band center, $\omega_r'/ 2\pi$ & 5.717\,GHz   \\
  \hline
 Neighbor coupling, $J/2\pi$ & 249\,MHz   \\
 \hline
 Next-Neighbor coupling, $J^{(2)}/2\pi$ & 38\,MHz   \\
 \hline
 Edges coupling   $\kappa/2\pi$ & 12\,MHz \\
  \hline
  Modes self Kerr, $K/2\pi$ & 0.1\,MHz   \\
 \hline
  Lattice constant , $d$ & 200\,$\mu$m \\
 \hline
  Band edge group index, $n_{g,e}$ & divergent \\
 \hline
   Band center group index, $n_{g,c}$ & 975 \\
 \hline
  \multicolumn{2}{c}{Qubits}\\
 \hline
\Qone maximum frequency $\omega_{q1,m}/2\pi$ & 6.322\,GHz  \\
 \hline
\Qtwo maximum frequency  $\omega_{q2,m}/2\pi$ & 6.606\,GHz  \\
 \hline
\Qone anharmonicity  $\beta_1/2\pi$ & -266\,MHz   \\
 \hline
\Qtwo anharmonicity  $\beta_2/2\pi$ & -257\,MHz\\
 \hline
\Qone-array coupling $g_1/2\pi$ & 338\,MHz   \\
 \hline
\Qtwo-array coupling   $g_2/2\pi$ & 311\,MHz   \\
 \hline
Nonradiative decay $\gamma_{q1,2}/2\pi$ & $\approx 50$\,kHz   \\
 \hline
\multicolumn{2}{c}{Read-out Resonators}\\
 \hline
 RO-\Qone frequency  $\omega_{ro1}/2\pi$ & 4.280\,GHz \\
 \hline
RO-\Qtwo frequency  $\omega_{ro2}/2\pi$ &  4.412\,GHz  \\
 \hline
 RO-\Qone coupling $h_1/2\pi$ & 98$\pm$1\,MHz \\
 \hline
RO-\Qtwo coupling   $h_2/2\pi$ & 89$\pm$1\,MHz \\
 \hline
Nonradiative decay   $\gamma_c/2\pi$ & $\approx 50$\,kHz   \\
\end{tabular}
\end{table}

\section{Theoretical model}\label{app:theory}
 
\subsection{Circuit model}\label{App.H}
\begin{figure}%
\includegraphics[width=\linewidth]{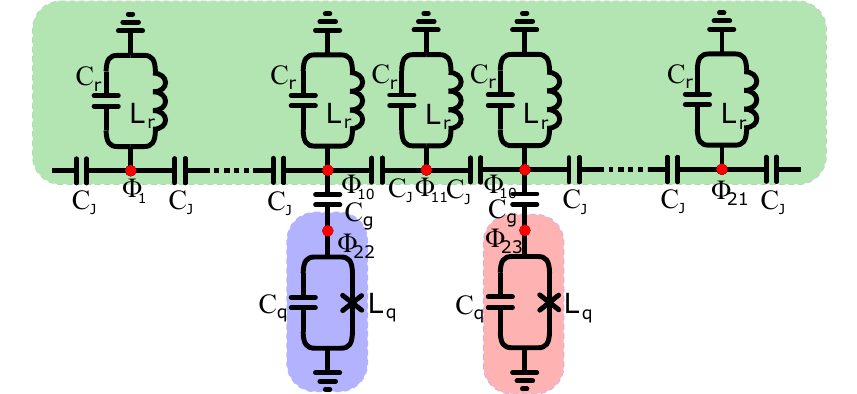}
\caption{Circuit model for $N=21$ capacitively coupled LC-resonators (in green). The resonators at the edges of the array are capacitively coupled to 50\,$\Omega$ transmission line. Two transmons qubits are coupled at sites $x=10$ (blue) and $x=12$ (red).}
\label{fig:electr_circuit}
\end{figure}
The ideal circuit model of the system is sketched in \fig \ref{fig:electr_circuit}  and shows $N$  LC-resonators, with capacitance $C_r$ and inductance $L_r$,
capacitively coupled in series via a capacitance $C_J$. Two sites of the array, $x_1=10$ and $x_2=12$, are coupled to two transmon qubits, represented by a Cooper pair box with capacitance $C_q$ and Josephson energy $E_J$, via  the coupling $C_{gi}$. 

To start, we consider the bare coupled cavity array waveguide, whose unit cell has a lattice constant $d=200\,\mu$m.
For probing frequency $\omega$ that respects the homogeneity condition $d<\lambda_g/4$ for the electromagnetic waves, we can treat the coupled-cavity array as a composite right-left-handed transmission line.

Defining the lattice unit impedance $Z_l(\omega)=1/i\omega C_J$ and admittance $Y_l(\omega)=(i\omega C_r+1/i\omega L_r)$, we can follow standard procedures \cite{CRLMeta} to  obtain the dispersion relation:

\begin{equation}    \label{eq:dispersion_electrical}
\cos (kd)=1+\frac{1}{2}Y_lZ_l.
\end{equation}
This equation can be recast in terms of the resonator  frequency, $\omega_r=1/\sqrt{\bar C_r L_r}$, with $\bar C_r=C_r+2C_J$, in the limit of $C_J\ll \bar C_r$  and reads
\begin{equation}\label{eq:dispersion_electrical2}
    \omega=\omega_r+C_J Z_r \omega_r^2 \cos (kd),
\end{equation}
where $Z_r=\sqrt{L_r/\bar C_r}$ is the characteristic impedance of the resonator. From the dispersion relation \eq \eqref{eq:dispersion_electrical}, we calculate the group velocity at the center of the band, $\omega=\omega_r$,  
\begin{equation}
    v_g(\omega_r)=C_JZ_r\omega_r^2 d,
\end{equation}
which gives us a left handed dispersion with negative phase velocity \cite{CRLMeta}.
Finally, we observe that the characteristic impedance of the transmission line is equal to: 
\begin{equation}
   Z(\omega)=\sqrt{\frac{Z_l}{Y_l}}\approx\sqrt{\frac{L_r}{C_J}}\sqrt{\frac{1}{1-\left(\frac{\omega}{\omega_r}\right)^2}}.
\end{equation}

After having characterized the waveguide properties we can take into consideration the full circuit model including the qubits. The Lagrangian of the circuit  can be written  in terms of the resonator and qubit fluxes, $\Phi_x$ and $\Phi_{qi}$ respectively \cite{devoret1995quantum}, and reads
\begin{equation}\label{lagrangian}
\mathcal{L}=\frac{1}{2}\mathbf{\dot \Phi}^T \mathcal{C}\mathbf{\dot\Phi}-V
\end{equation}
where $\mathbf{\dot \Phi}^T=(\dot\Phi_{q1}, \dot\Phi_{q2}, \dot\Phi_{1},...,  \dot\Phi_{N})$ and
\begin{equation}
V=\sum_x \frac{\Phi_x^2}{2L_x}-\sum_{i=1}^2 E_J\cos{\left(\frac{\Phi_{qi}}{\Phi_0}\right)}
\end{equation}
is the potential energy term with $\Phi_0=\hbar/(2e)$ being  the flux quantum. 
The first term of \eqref{lagrangian} is the kinetic term that is governed by the 
capacitance matrix

\begin{equation}
\mathcal{C}=\left(
\begin{smallmatrix}
 \bar C_{q1} &  0 & 0& 0&  \dots & -C_{g1} & 0 & 0 & \dots  & 0\\
0 &  \bar C_{q2}  & 0 & 0& & 0 & 0 & -C_{g2} & & 0 \\
0 &  0  &   \bar C_r & -C_J  &  &  0 & 0&  0 &  & 0 \\
0 &  0  & -C_J  &  \bar C_r   & \ddots  &  0 & 0&  0 & & 0 \\
\vdots &     &  & \ddots & \ddots &   &   &   &  & \vdots\\
-C_{g1} &  0  &  0 & 0 & &   \bar C_r  & -C_J & 0 & & 0 \\
0 &  0  & 0&  0 &  &  -C_J & \bar C_r &  -C_J  &  & 0\\
0 &   -C_{g2}  & 0 & 0 & &  0  & -C_J &\bar C_r  & \ddots & 0 \\
 \vdots &   &   & & &   &   &   \ddots & \ddots &  -C_J \\
 0 &  0  & 0  &  0 & \dots  &  0 & 0&  0& -C_J  & \bar C_r  \\
\end{smallmatrix}\right)
\end{equation}
where  $\bar C_{qi}=C_{qi}+C_{gi}$.
The Hamiltonian of the system is obtained from the Lagrangian \eqref{lagrangian} via the usual Legendre transformation \cite{devoret1995quantum} $H=\mathbf{Q}^T\mathbf{\dot\Phi}-\mathcal{L}$, where we introduced   the conjugate charge variables $\mathbf{Q}=\partial\mathcal{L}/\partial\mathbf{\dot\Phi}=\mathcal{C}\mathbf{\dot\Phi}$, with $\mathbf{Q}^T=(Q_{q1}, Q_{q2}, Q_{1},...,  Q_{N})$. In the matrix notation the Hamiltonian reads
\begin{equation}\label{H_charges}
H=\frac{1}{2}\mathbf{Q}^T \mathcal{C}^{-1}\mathbf{Q}+V
\end{equation}
where now the coupling between the charges are governed by the inverse of the capacitance  matrix $\mathcal{C}^{-1}$. This matrix when inverted connects all the elements of the circuit. Nevertheless, we can use the fact that the coupling capacitances are much smaller than the resonator ones, $C_J, C_{gi}\ll\bar C_r,\bar C_{qi}$, to obtain the following simplified expression
\begin{equation}
\frac{1}{\mathcal{C}}=\left(
\begin{smallmatrix}
  \frac{1}{\bar C_{q1}} &  0 & 0& 0&  \dots & \frac{1}{\bar C_{g1}}& 0 & 0 & \dots  & 0\\
0 &   \frac{1}{\bar C_{q2}}   & 0 & 0& & 0 & 0 & \frac{1}{\bar C_{g2}}& & 0 \\
0 &  0  &   \frac{1}{\bar C_r}  & \frac{1}{\bar C_J}  &  &  0 & 0&  0 &  & 0 \\
0 &  0  &  \frac{1}{\bar C_J}  & \frac{1}{\bar C_r}  & \ddots  &  0 & 0&  0 & & 0 \\
\vdots &     &  & \ddots & \ddots &   &   &   &  & \vdots\\
 \frac{1}{\bar C_{g1}} &  0  &  0 & 0 & &   \frac{1}{\bar C_r} &\frac{1}{\bar C_J}& 0 & & 0 \\
0 &  0  & 0&  0 &  &  \frac{1}{\bar C_J}&  \frac{1}{\bar C_r} &  \frac{1}{\bar C_J}&  & 0\\
0 &   \frac{1}{\bar C_{g2}}  &0 & 0 & &  0  & \frac{1}{\bar C_J} & \frac{1}{\bar C_r} & \ddots & 0 \\
 \vdots &   &   & & &   &   &   \ddots & \ddots &  \frac{1}{\bar C_J} \\
 0 &  0  & 0  &  0 & \dots  &  0 & 0&  0& \frac{1}{\bar C_J} &  \frac{1}{\bar C_r}  \\
\end{smallmatrix} \right)
\end{equation}
with the definitions   $\frac{1}{\bar C_{gi}}= \frac{C_{gi}}{\bar C_{qi}\bar C_r}$ and $\frac{1}{\bar C_J}= \frac{C_J}{\bar C^2_r}$.
In order to quantize the Hamiltonian \eq \eqref{H_charges} we first  express the charge and the flux of the resonators in terms of the  annihilation and creation operators:
\begin{equation}
\begin{split}
&Q_{x}=\sqrt{\hbar \bar C_{r}\omega_{r}/2}(a^\dagger+a)\\
&\Phi_{x}=i\sqrt{\hbar/(2 \bar C_{r}\omega_{r})}(a^\dagger-a).
\end{split}
\end{equation}

In this way the canonical commutation relations $[\Phi_xQ_{x'}]=i\hbar\delta_{x,x'}$ are satisfied.

The qubits can be described by nonlinear resonators with a Kerr nonlinearity $\beta_i$ and annihilation (creation) operators $b$ ($b^\dagger$). 
The qubit frequency is given by the frequency difference between the two lowest energy states, $|0\rangle$ and $|1\rangle$, of the transmon and it is a function of the flux on the qubit, i.e.  $\omega_{qi}=\omega_{qi}(\Phi_{qi})$. The transition between the first two transmon levels is determined by the dipole moment $D_{qi}=\langle 1|Q_{qi}|0\rangle$ and defines the qubit-cavity couplings $g_i=D_{qi}\sqrt{\bar C_r\omega_r/(2\bar C^2_g)}$. Finally, by defining the cavity-cavity hopping $J=\bar C_r\omega_r/(2\bar C_J)$ we obtain the system Hamiltonian given in \eq \eqref{eq:sys_hamiltonian}.
Note that in this derivation we neglected direct parasitic capacitive couplings, disorder in the circuit elements and we made some simplifications on the inverse capacitance matrix. We discuss the effect of these approximations in detail in \app \ref{App. Imperfections}.

\subsection{Bare coupled-cavity array in the tight binding picture}\label{App.naked_array}
Here we re-discuss the bare array of coupled resonators starting from the standard tight-binding model with uniform nearest-neighbor couplings
	\begin{equation}\label{H_CCA}
		H_{\rm CCA}=\omega_r\sum_{x=1}^N a_x^{\dagger} a_x+J\sum_{x=1}^{N-1}\left(a^{\dagger}_{x+1}a_x+\rm H.c \right)\,
	\end{equation}
	with ladder operators $a_x$ fulfilling usual bosonic commutation rules, $[a_x,a^\dag_{x'}]=\delta_{x,x'}$.
	It is worth stressing that here $N$ must be finite and the array subject to open boundary conditions, in contrast to most treatments in the literature of bound states which focus on the thermodynamic limit and periodic boundary conditions.
	
	The array's free Hamiltonian \eqref{H_CCA} is diagonalized as (see e.g.~Ref.~\cite{Ciccarello2011})
	\begin{equation}
		H_{\rm CCA}=\sum_{k}\omega_k a_k^{\dagger} a_k\,,
	\end{equation}
	with the normal modes $ a_k$ given by 
	\begin{equation}\label{normal}
		 a_k=\sqrt{\frac{2}{N+1}}\,\sum_{x=1}^N\sin{(kx)} a_x
	\end{equation}
	and the normal frequencies $\omega_k$ by
	\begin{equation}
		\label{eq:dispersion}
		\omega_k=\omega_r+2J\cos k
	\end{equation}
	with 
	\begin{equation}
		k=\frac{m\pi}{N+1}\;\;\,\,\,(m=1,2,...,N)\,.
	\end{equation}
	Note that the discrete spectrum \eqref{eq:dispersion} obtained by a  tight binding array with open boundary conditions coincides with the one obtained by the circuit model in Eq. \eqref{eq:dispersion_electrical2}.


\subsection{One-atom bound states}\label{App.BS1atom1ex}

Consider the case that only one atom is effectively coupled to the coupled-cavity array (the other being far-detuned from the photonic band). Then the total Hamiltonian \eq \eqref{eq:sys_hamiltonian} in terms of normal modes \eqref{normal} and in a frame rotating at frequency $\omega_r$, reads (we omitted the label $i$ in this case)
\begin{align}\label{Htotalk}
 H=\,\,& 2J \sum_{k}\cos k\,   a^{\dagger}_k a_k  +  \delta  \,b^\dagger  b\\
&+g\,\sqrt{\frac{2}{N+1}}\,\sum_{k}\sin(k x_q )\left( a_k^{\dagger} b + a_k b ^{\dagger}\right)\,,
 \end{align}
where $\delta=\omega_q-\omega_r$ is the detuning of the qubit from the bare frequency of each resonator and $x_q$ the cavity to which the atom is directly coupled. Note the sine-shaped atom-mode interaction strength (colored coupling), stemming from the open boundary conditions which the array is subject to~\cite{Longhi}. As shown next, such a colored coupling generally could lead to results slightly different from the usual white coupling under periodic boundary conditions \cite{Lombardo2014,Longhi,calajoBS,Tao}.

Within the single-excitation subspace defined by $n=1$, an atom-photon bound state $\ket{\phi}$ can be worked out as an eigenstate of the total Hamiltonian whose corresponding energy lies outside the photonic band, that is such that $H\ket{\phi}=\hbar\omega\ket{\phi}$ with $|\omega|>2J$. Expanding $\ket{\phi}$ in the basis $\{b^\dagger|g,0\rangle,\,\{a_k^{\dagger}|g,0\rangle\}\}$ as
\begin{equation}
|\phi\rangle=\left(c_q\, b^\dagger +\sum_k u_k a_k^{\dagger}\right)|g,0\rangle\,
\end{equation}
and then inserting this into the Schr\"odinger equation yields that the frequency $\omega$ must fulfills the equation
 \begin{equation}\label{EI0}
\hbar(\omega-\delta)=\Sigma(\omega)\,,
\end{equation}
where the self-energy $\Sigma(\omega)$  is given by 
 \begin{equation}\label{self}
 \begin{split}
\Sigma(\omega)/\hbar=\,\,&\frac{2g^2}{N+1}\sum_k \frac{\sin^2{\left(\frac{k}{2}x_q\right)}}{\omega-2J\cos \frac{k}{2}}\\
=\,\,&\frac{g^2}{2J}\,\,\frac{1-e^{-2\frac{x_q}{\lambda}}-e^{-2\frac{N+1-x_q}{\lambda}}+e^{-2\frac{N+1}{\lambda}}}{\sinh\left[{\frac{1}{\lambda}\left(1-e^{-2\frac{N+1}{\lambda}}\right)}\right]}\,
\end{split}
\end{equation}
with
 \begin{equation}\label{lambda}
\lambda(\omega)=\left({\arccosh{\tfrac{|\omega|}{2J}}}\right)^{-1}\,\,.
\end{equation} 
The solutions of Eq.\eqref{EI0} with $|\omega|>2J$
correspond to the atom-photon bound states.
 In the standard case of periodic boundary conditions, two bound states -- one with energy above and one below the band -- always exist \cite{calajoBS,Tao}. In our case, however, the presence of array edges may affect the existence of bound states. By substituting in the eigenvalue \eq\eqref{EI0} the limiting values of the self-energy $\Sigma(\omega)$ for $\omega\rightarrow \pm 2J$, the  condition for existence of bound states is obtained as
\begin{equation}\label{existence_cond}
g^2>\frac{J(N+1)(2J\mp\delta)}{x_q(N+1-x_q)}\,,
\end{equation}
where $-$ ($+$) indicates a solution with energy above (below) the band. 
In our experimental setup, $g\sim J$, $N=21$ and $x_q=10,12$ so that Eq.~\eqref{existence_cond} is always satisfied and the self-energy is well approximated by performing the thermodynamic limit
 \begin{equation}\label{self_cont}
\Sigma(\omega)/\hbar\simeq \frac{g^2}{\omega\sqrt{1-\frac{4J^2}{\omega^2}}}\,.
\end{equation}
In practice, this means that the atom in our setup is sufficiently far from the edges and the number of resonators large enough that the bound state can be calculated as if the array were infinitely long (in line with standard treatments).

Following Ref.~\cite{calajoBS}, the bound state corresponding to a solution $\omega_{\rm BS}$  can thus be worked out in the form
 \begin{equation}\label{boundstate_form}
 \begin{split}
|\phi\rangle=& 
\left[ \cos{\theta} \, b^\dagger +(-1)^s  \sin{\theta}   \,\alpha^\dag \right] |g,0\rangle\,.
\end{split}
\end{equation} 
where $\theta$ is given by
\begin{equation}\label{theta}
\cos  \theta=\left(1+\frac{g^2}{\omega_{\rm BS}^2\left(1-\tfrac{4J^2}{\omega_{\rm BS}^2}\right)^{\tfrac{3}{2}}}\right)^{-\tfrac{1}{2}}\mbox{,}
\end{equation} 
while 
\begin{equation}\label{alop}
	\alpha =   \sum_x\,\frac{ s^{|x-x_q|} \,e^{-\frac{|x-x_q|}{\lambda_\pm}} }{{\sqrt{\coth{\frac{1}{\lambda_\pm}}}}} \, \,a_{x}
\end{equation}
defines a bosonic ladder operator. Here, $\lambda=\lambda(\omega_{\rm BS})$ as given by \eqref{lambda}, while $s={\rm sgn}(\omega_{\rm BS})$ so that $s=+1$ when $\omega_{\rm BS}$ lies above the band and $s=-1$ when it falls below.

The mixing angle $\theta$ measures the degree of hybridization: the dressed state is fully atomic for $\theta=0$ and fully photonic for $\theta=\pi/2$. Eq.~\eqref{alop} fully defines the (normalized) photonic component, showing that the spatial mode is exponentially localized around the atom's position $x_q$. 
Accordingly, parameter $\lambda$ represents the localization length of the photonic cloud surrounding the atom.

\subsection{Two-atom bound states}\label{App.2atom}

Let us now study the case of bound states in the single-excitation sector when two atoms are  coupled to resonators $x_1$ and $x_2$ respectively.
The total Hamiltonian is the natural generalization of \eqref{Htotalk} and will now feature the detuning of each qubit $\delta_i=\omega_{qi}-\omega_r$ and the corresponding coupling strength $g_i$ with $i=1,2$.

In this case, an effective interaction between bound states arises when the photonic clouds of the individual bound states overlap. This interaction 
changes the bound-state energies, which are now given by the real solutions with $|\omega|>2J$ of the transcendental equation~\cite{calajoBS} 
\begin{equation}\label{eig2e}
\begin{split}
\left(\omega-\delta_1-\Sigma_1\right)\left(\omega-\delta_2-\Sigma_2\right)=\Sigma_1\Sigma_2\,e^{-2\frac{|x_1- x_2|}{\lambda}}\,.
\end{split}
\end{equation}
Here, $\Sigma_i$ is the self-energy of the $i$th qubit (in absence of the other qubit) as given by Eq.~\eqref{self_cont} for $g=g_i$.

Eq.~\eqref{eig2e} admits up to four real solutions (and as many bound states).
Based on the performed measurements in our experiment, we  focus on the case of bound-state energies above the band. Then the pair of bound states with energies $\omega_\pm$ are given by 
\begin{equation}\label{2state}
 |\phi_{\pm}\rangle = \tfrac{1}{\sqrt{1+|\xi|^2}} \left[ D_{\pm}^\dag(x_1)  \pm \xi D_{\pm}^\dag(x_2)\right] |g_1,g_2,0\rangle,
 \end{equation}
where we defined the dressed ladder operators 
 \begin{equation}\label{2atoper}
 D_{\pm}(x_i) = \cos(\theta_\pm) b_i + \sin(\theta_\pm) \tfrac{1}{\mathcal{N}_{\pm}}\sum_x e^{-\frac{|x-x_i|}{\lambda_\pm}} a_x,
 \end{equation}
and the mixing angle
 \begin{equation}\label{2theta}
\cos{\theta_\pm}=\left(1+\frac{g_1g_2\mathcal{N}^2_{\pm}}{4J^2\sinh^2{\frac{1}{\lambda_\pm}}}\right)^{-\frac{1}{2}}
\end{equation}
with
  \begin{equation}
 \mathcal{N}_{\pm}=\sqrt{\coth{\tfrac{1}{ \lambda_\pm}}\left(1\pm e^{-\frac{|x_1-x_2|}{\lambda_\pm}}\right)\pm |x_1-x_2|e^{-\frac{|x_1-x_2|}{\lambda_\pm}}}\,.\nonumber
 \end{equation}
 The parameter $\xi=\xi(\omega_{\pm})$ gives the amount of the hybridization between the two atoms and reads
 \begin{equation}
\xi(\omega_{\pm})=\frac{\sqrt{\Sigma_1\Sigma_2}e^{-\frac{|x_1-x_2|}{\lambda}}}{\omega_{\pm}-\delta_2-\Sigma_2}.
 \end{equation}
 In the regime of equal coupling $g_1=g_2$ (quite close to our experimental realization) and in correspondence with the avoided crossing, $\delta_1=\delta_2$, the two bound states get completely hybridized,  $\xi(\omega_{\pm})=1$, and the labels  $+(-)$ signify the bound states with even (odd) symmetry with respect to the atoms' midpoint.


Under certain conditions, the interaction can push the odd-parity bound state frequency into the propagating band (in which case the state just no longer exists). Similarly to \eqref{existence_cond}, an existence condition for this anti-symmetric state can be worked out by taking the limit $\omega\rightarrow  2J^+$ in \eqref{eig2e}, which yields (for $g_1=g_2=g$)
 \begin{equation}\label{melting_cond}
g^2>\frac{J(4J-\delta_1-\delta_2)}{|x_1-x_2|}.
\end{equation}
The melting of this bound state into the photonic band reported in \fig \ref{fig:interaction} indeed occurs for parameters such that \eqref{melting_cond} is not satisfied.

The interaction between one-atom bound states can cause also a (coherent) excitation transfer between the two atoms described by an effective spin Hamiltonian (the field is adiabatically eliminated). This occurs in the dispersive regime where the qubits are far-detuned from the band edge, i.e., $\omega_{qi}-\omega_r-2J\gg g_i$, and $\cos\theta_{\pm}\approx 1$ such that the bound states are mostly atomic. The effective spin Hamiltonian then reads \cite{Efi2013,Douglas2015}
\begin{equation}\label{eq.effH}
H_{\rm eff}/\hbar=\frac{g_1g_2}{\delta_e}\sum_{ij=1}^2e^{-\frac{|x_i-x_j|}{\lambda}}b_i^\dagger b_j\,,
\end{equation}
where we assumed that the atoms are tuned on resonance with one another, i.e., $\omega_{q1}=\omega_{q2}=\omega_q$, and set $\delta_e=\omega_q-\omega_r-2J$ (detuning from the band edge). This Hamiltonian captures the excitation exchange dynamics between the qubits in \fig\ref{fig:time_interaction}. The coherence time is just given by the single atom bound state decay into the waveguide and into the other dissipation channels as discussed in the main text.

\subsection{Two-photon bound states}\label{App.2BS}

We next study the energy of two-excitation bound states. In the ideal case $\beta\rightarrow-\infty$ (ideal two-level atom case), two-photon bound states are known to occur entailing strong nonlinear effects \cite{calajoBS,Tao}. Yet, even for finite $\beta$, the nonlinear energy spacing of levels above the first two can still affect two-excitation bound states causing measurable deviations from the fully linear case $\beta=0$ (as discussed in the main text). To show this in more detail, we first note that any state in the two-excitation sector can be written as
\begin{equation}\label{eq:2phoansatz}
	\begin{split}
		|\phi^{(2)}\rangle=&\left(\frac{1}{\sqrt{2}}\sum_{i,j=1}^2c_{ij}b_i^\dagger b_j^\dagger+\sum_{i=1}^2\sum_x c_i(x)b_i^\dagger a_x^{\dagger}\right.\\
		&\left.+\frac{1}{\sqrt{2}}\sum_{x,y} u(x,y)a_{x}^{\dagger}a_{y}^{\dagger}\right)|g,0\rangle\,.
	\end{split}\,.
\end{equation}


Here, $c_{ij}$ is the probability amplitude of having one excitation on transmon $i$ and one on transmon $j$ (including the case $i=j$), while $c_{i}(x)$ is the probability amplitude corresponding to one excitation on the $i$-th transmon and another one in the waveguide. Finally, $u(x,y)=u(y,x)$ is the symmetric wavefunction of the two-photon bound state component. By plugging \eq \eqref{eq:2phoansatz} in the Schr\"odinger equation generated by Hamiltonian \eq \eqref{eq:sys_hamiltonian}, we numerically solve the resulting set of coupled equations. 

In the following, we separately address the one- and two-atom case.

\subsubsection{One atom}


\begin{figure}%
	\includegraphics[width=\linewidth]{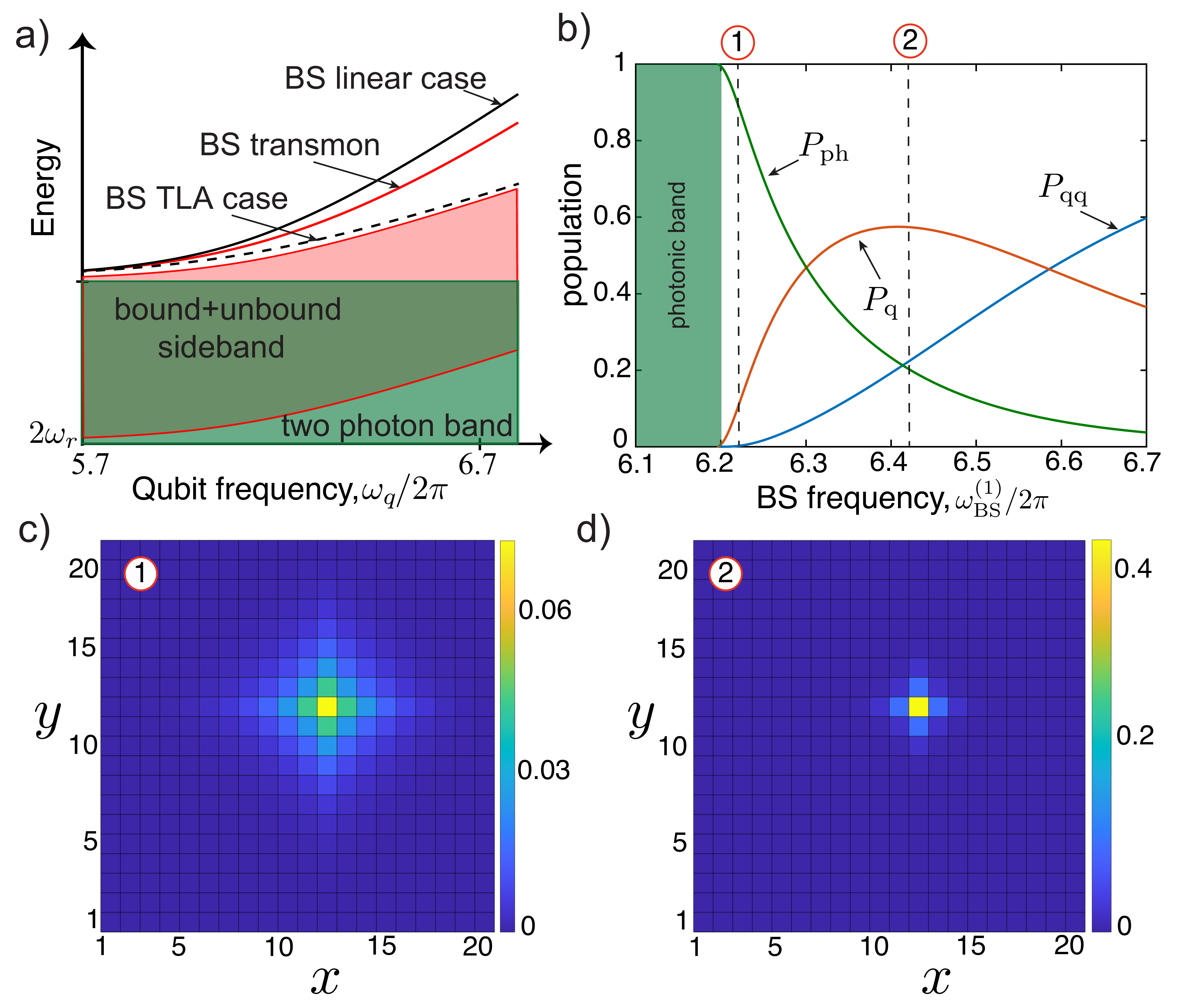}
	\caption{(a)Sketch of the two-excitation spectrum (upper part) for a single qubit as function of the qubit frequency. The bound state in the case of a linear resonator (black solid line) and an ideal two-level qubit (black dashed lines) are also shown for comparison (see main text for more details). (b) Excitation distribution of the two-photon bound state as a function of the single-excitation bound state frequency. Specifically, we plot the probability to find one or two excitations on the transmon, respectively given by $P_{q}=\sum_{x=1}^N|c_{2}(x)|^2$ and $P_{qq}=|c_{22}|^2$, along with the two-photon population $P_{ph}=\sum_{x=1}^N\sum_{y=1}^N|u(x,y)|^2$. (c)-(d) Spatial profile of the normalized two-photon wavefunction for the frequencies highlighted by the dashed vertical line in panel (b). The considered parameters are those corresponding to \Qtwo\, (see Table \ref{tbl:parameters}) with $\omega_r/2\pi=5.7$ GHz, $J/2\pi=249$ MHz, $g_{2}/2\pi=311$ MHz and $\beta_{2}/2\pi=-257$ MHz for the transmon qubit.}.
	\label{fig:2ex_BS}
\end{figure}

\fig\ref{fig:2ex_BS}(a) shows the upper part of the two-excitation spectrum, $\omega>2\omega_r$, as a function of the atom frequency. The full spectrum (of which only the upper part is shown) features a band of two-photon unbound states defined by $\omega\in[2(\omega_r-2J),2(\omega_r+2J)]$ plus a pair of sidebands $\omega\in[\omega^{(1)}_{\rm BS}-2J,\omega^{(1)}_{\rm BS}+2J]$, where $\omega^{(1)}_{\rm BS}$ is the energy of the single-excitation bound states. Additionally, there exist a pair of two-photon bound states (see section \ref{Sec.twoex}) with discrete energies $\omega^{(2)}_{\rm BS}$ such that $|\omega^{(2)}_{\rm BS}-\omega_r|>4J$ (one above and one below the continuous bands).  For comparison, we also plot the bound state energies in the limiting cases of a linear resonator (black solid line) and a two-level atom (black dashed line). In  \fig\ref{fig:BS_Anharmonicity} of the main text we defined the dressed-state anharmonicity as the difference in the bound state energy between the linear and nonlinear case.
While \fig\ref{fig:2ex_BS}(a) shows occurrence of a two-excitation bound state, it is natural to wonder how hybridized such a state is, and additionally, if it features a significant two-photon component (or alternatively if it is mostly populated by the excitation of the second transmon level). To clarify this point, in \fig\ref{fig:2ex_BS}(b) we plot the population distribution of the two-excitation bound state, which clearly shows the photon dressed nature of the bound state in the considered parameter regime. Notably, the two-photon wavefunction exhibits a different localization length depending on the qubit frequency, in this respect similarly to the single-excitation case [see \fig\ref{fig:2ex_BS}(c)-(d)].

\subsubsection{Two atoms}

\begin{figure}%
	\includegraphics[width=\linewidth]{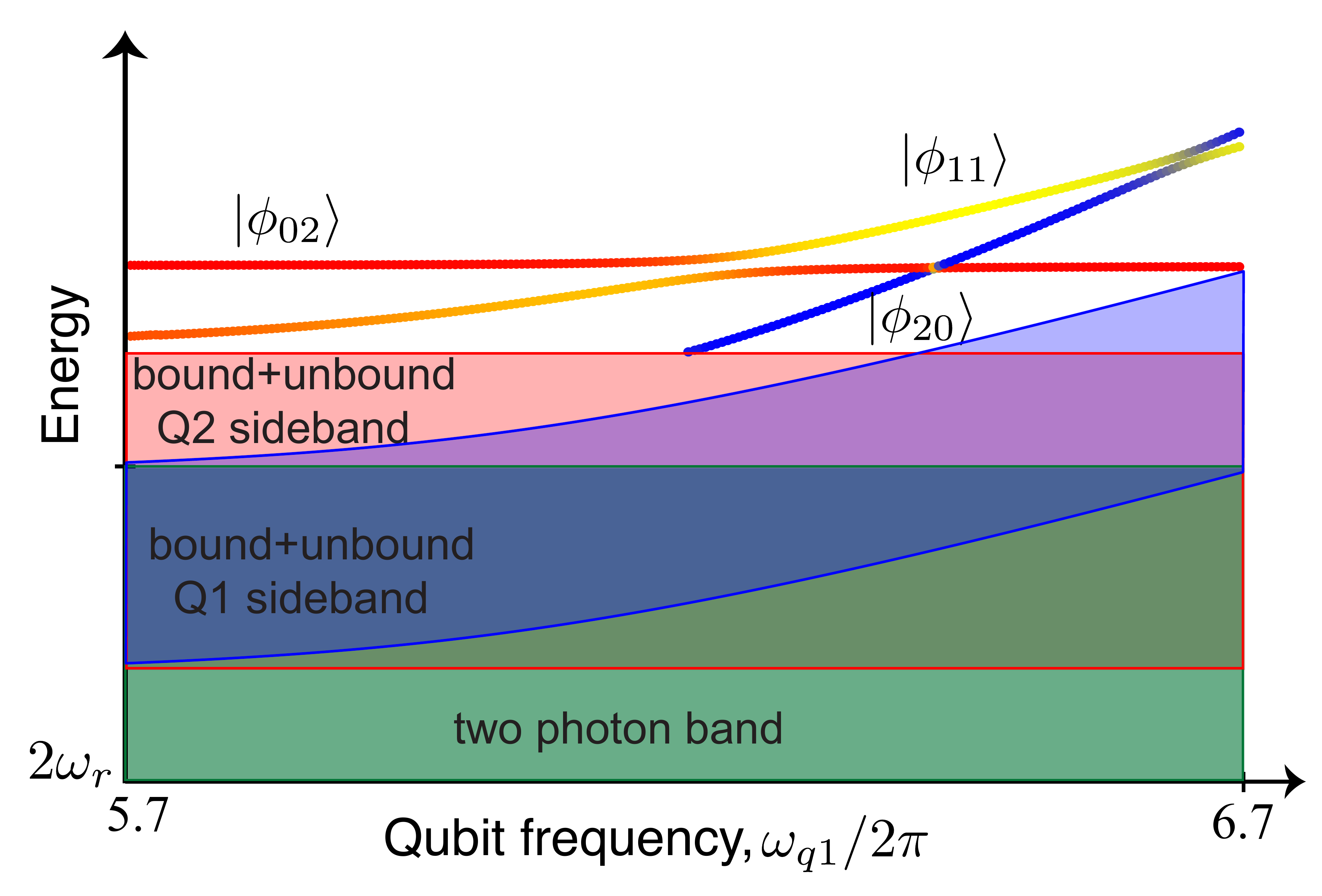}
	\caption{Sketch of the two-excitation spectrum (upper part) as a function of the \Qone\, frequency in the case of two qubits. We set the frequency of Q$_2$ to $\omega_{q2}/2\pi=6.45$GHz with the remaining parameters fixed to   $\omega_r/2\pi=5.7$ GHz, $J/2\pi=249$ MHz, $g_{2}/2\pi=311$ MHz, $g_{1}/2\pi=338$ MHz, $\beta_{2}/2\pi=-257$ MHz $\beta_{1}/2\pi=-266$ MHz. In the way the bound states correspond to the one shown in \fig\ref{fig:cross_kerr}.}
	\label{fig:2ex_2at_BS}
\end{figure}

As discussed in the main text, the two-excitation spectrum in the case of two atoms becomes quite involved, as shown in \fig\ref{fig:2ex_2at_BS}. Besides the 
two-photon unbound states with energies $\omega\in[2(\omega_r-2J),2(\omega_r+2J)]$ (green box), two sidebands occur (above the main band) with energies $\omega\in[\omega^{(1)}_{\rm +}-2J,\omega^{(1)}_{\rm +}+2J]$ and $\omega\in[\omega^{(1)}_{\rm -}-2J,\omega^{(1)}_{\rm -}+2J]$, where the $\pm$ sign refers to the two-atom single-excitation bound state discussed in Section \ref{App.2atom}. Above the bands, there appear three bound states stemming from the hybridization of the bare qubit states $|20\rangle,\,|02\rangle$ and $|11\rangle$. Once coupled to the array, the transmons get highly hybridized both with one another and the array field, resulting in the dressed states $|\phi_{20}\rangle,\,|\phi_{02}\rangle$ and $|\phi_{11}\rangle$ (the labels reflect the resemblance of each dressed state to the bare state of corresponding indices away from the avoided crossings). In particular, the color scale used in \fig\ref{fig:2ex_2at_BS} reflects the population distribution among the two qubits, ranging from blue (excitation on \Qone) to red (\Qtwo)



\subsection{Array Transmission}\label{app:input-output}

We calculate the transmission through the array applying the input-output theory relation to the Heisenberg equation \cite{walls_quantum_2008}. The intra-cavities field $a_x$ for the $x$-th resonator, and $b_m$ for the $m$-th qubit, with a driving field $a_{\rm in}(t)=a_{\rm in}e^{-i\omega t}$ applied on resonator $x=1$
\begin{align}
    &\dot a_x(t)=i\left[H, a_x(t)\right]-\frac{1}{2}\kappa a_x(t)+\sqrt{\kappa_{\rm r}}\delta_{x 1}a_{\rm in}\\
     &  \dot b_m(t)=i\left[H, b_i(t)\right]-\frac{1}{2}\gamma b_m(t)
\end{align}
where $\kappa=\delta_{x 1}\kappa_{\rm r}+ \delta_{x 21}\kappa_{\rm r}+\kappa_{\rm nr}$ is the sum of radiative (in case of the edges cavities) and nonradiative decay rate for the resonators, $\gamma$ is the decay rate for the qubits, and $H$ is the Hamiltonian in \eq \eqref{eq:sys_hamiltonian}. In the steady state, $a_x(t)=e^{-i\omega t}a_x$ and the equations for the resonators and qubits fields become
\begin{align}
\begin{split}
   0= & \left( \Delta_{\rm r}-\frac{i}{2}\kappa \right)a_x+J(a_{x+1}+a_{x-1})\\
    &+ \delta_{xm}gb_m+i\sqrt{\kappa_{\rm r}}\delta_{x 1}a_{\rm in}
\end{split}
\\
  0= &\left( \Delta_{m}+\frac{i}{2}\gamma \right)b_m+\delta_{xm}ga_x
\end{align}
where we defined the tuning $\Delta_{\rm r}=\omega_r-\omega$ and $\Delta_{ m}=\omega_{qi}-\omega$. Solving the algebraic system and applying the input-output relation $\langle a_{out}\rangle+\langle a_{in}\rangle=\sqrt{\kappa_{\rm nr}}\langle a_x\rangle$ we calculate the transmission coefficient from cavity 1 to 21.


\section{Experimental imperfections}\label{App. Imperfections}

\subsection{Parasitic capacitance}
\label{app:parassitic}
As observed in the main text, the frequency distribution of the coupled-cavity array modes and the BSs interaction strength, cannot be quantitatively reproduced neglecting capacitive couplings beyond the nearest neighbor.

Solving Poisson's equation with a finite element method solver (Comsol Multiphysics, electrostatic package) we estimate a parasitic capacitance between next nearest neighbor resonators $C^{(2)}_J\approx 0.52$\,fF and the one between qubits and resonator $C^{(2)}_{g}\approx 0.73$\,fF.  Adding this contribution to the capacitance matrix $\mathcal{C}$, we obtain
\begin{equation}
\left(
\begin{smallmatrix}
 \bar C_{q1} &  0 & 0& 0&  \dots & -C^{(1)}_{g1} & -C^{(2)}_{g} & 0 & \dots  & 0\\
0 &  \bar C_{q2}  & 0 & 0& & 0 & -C^{(2)}_{g} & -C^{(1)}_{g2} & & 0 \\
0 &  0  &   \bar C_r & -C_J  &  &  0 & 0&  0 &  & 0 \\
0 &  0  & -C_J  &  \bar C_r   & \ddots  &  0 & 0&  0 & & 0 \\
\vdots &     &  & \ddots & \ddots &   &   &   &  & \vdots\\
-C_{g1} &  0  &  0 & 0 & &   \bar C_r  & -C_J & -C^{(2)}_J & & 0 \\
 -C^{(2)}_{g}  &  -C^{(2)}_{g}   & 0&  0 &  &  -C_J & \bar C_r &  -C_J  &  & 0\\
0&   -C_{g2}  & 0 & 0 & &  -C^{(2)}_J  & -C_J &\bar C_r  & \ddots & 0 \\
 \vdots &   &   & & &   &   &   \ddots & \ddots &  \vdots \\
 0 &  0  & 0  &  0 & \dots  &  0 & 0&  0& -C_J  & \bar C_r  \\
\end{smallmatrix}\right)
\end{equation}
The inverse capacitance matrix, retaining only the first order terms in $C_{gi}$, $C_J$, $C_J$, $C_J^{(2)}$ and $C_{gi}^{(2)}$ becomes
\begin{equation}
\left(
\begin{smallmatrix}
  \frac{1}{\bar C_{q1}} &  0 & 0& 0&  \dots & \frac{1}{\bar C_{g1}}& \frac{1}{\bar C^{(2)}_{g1}} & 0 & \dots  & 0\\
0 &   \frac{1}{\bar C_{q2}}   & 0 & 0& & 0 & \frac{1}{\bar C^{(2)}_{g1}} & \frac{1}{\bar C_{g2}}& & 0 \\
0 &  0  &   \frac{1}{\bar C_r}  & \frac{1}{\bar C_J}  &  &  0 & 0&  0 &  & 0 \\
0 &  0  &  \frac{1}{\bar C_J}  & \frac{1}{\bar C_r}  & \ddots  &  0 & 0&  0 & & 0 \\
\vdots &     &  & \ddots & \ddots &   &   &   &  & \vdots\\
 \frac{1}{\bar C_{g1}} &  0  &  0 & 0 & &   \frac{1}{\bar C_r} &\frac{1}{\bar C_J}& \frac{1}{\bar C^{(2)}_J} & & 0 \\
\frac{1}{\bar C^{(2)}_{g1}} &  0  & 0&  0 &  &  \frac{1}{\bar C_J}&  \frac{1}{\bar C_r} &  \frac{1}{\bar C_J}&  & 0\\
\frac{1}{\bar C^{(2)}_{g2}} &   \frac{1}{\bar C_{g2}}  & 0 & 0 & &   \frac{1}{\bar C^{(2)}_J}  & \frac{1}{\bar C_J} & \frac{1}{\bar C_r} & \ddots & 0 \\
 \vdots &   &   & & &   &   &   \ddots & \ddots &  \vdots \\
 0 &  0  & 0  &  0 & \dots  &  0 & 0&  0& \frac{1}{\bar C_J} &  \frac{1}{\bar C_r}  \\
\end{smallmatrix} \right)
\end{equation}
with the definitions $\frac{1}{\bar C^{(2)}_J}= \frac{C^{(2)}_J}{\bar C^2_r}$ and $\frac{1}{\bar C^{(2)}_{gi}}=\frac{C^{(2)}_{gi}}{\bar C_rC_{qi}}+\frac{C_r}{\bar C_{qi} \bar C_J}$. 

It is finally possible to write the Hamiltonian: 

\begin{equation}
\label{eq:sys_hamiltonian_full}
\begin{split}
    H/\hbar= & \sum_{x=1}^{N}\omega_r a_x^\dagger  a_x
    +\sum_{x=1}^{N-1}J\left( a_x^\dagger  a_{x+1}+ a_{x+1}^\dagger  a_{x}\right)\\
    +& J^{(2)}\sum_{x=1}^{N-2} \left(a^{\dagger}_{x+2}a_x+\rm H.c \right)\\ 
    + & \sum_{i=1}^{2}\omega_{qi} b_i^\dagger  b_{i}+\frac{1}{2}\beta_i b_i^\dagger b_i^\dagger b_i  b_i+g_i\left( a^\dagger_{x_i} b_i+ b_i^\dagger  a_{x_i}\right)\\
    +&\sum_{i=1}^2 g_i^{(2)}\left( a^\dagger_{x_i+1} b_i+ b_i^\dagger  a_{x_i+1}+a^\dagger_{x_i-1} b_i+ b_i^\dagger  a_{x_i-1}\right)\,
    \end{split}
\end{equation}
where we introduced the next neighbor couplings $g_i^{(2)}=D_{qi}\sqrt{\bar C_r\omega_r/\left[2  \left(\bar C^{(2)}_{gi}\right)^2\right]} $ and $J^{(2)}=\omega_r \bar C_r/(2 \bar C_J^{(2)})$.

\subsection{Interaction strength optimization}
\label{app:optimization}
\begin{figure}%
\includegraphics[width=\linewidth]{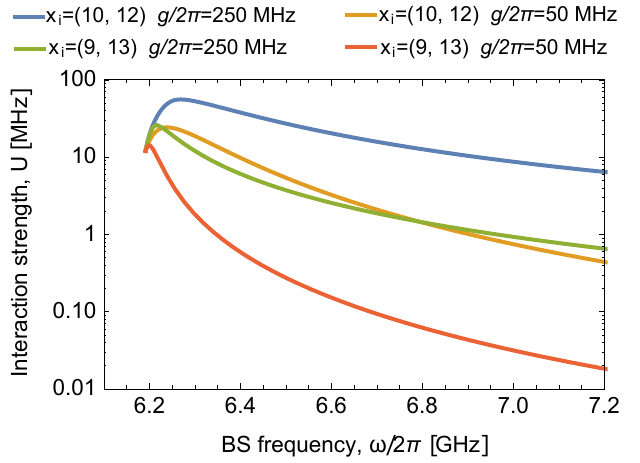}
\caption{\label{fig:optimization}Interaction strength between two bound states. The four lines represent the calculated coupling between two bound states, for qubits coupled to different array sites($x_i$) and with different qubit-cavity coupling strengths $g$. 
}
\end{figure}
In \fig \ref{fig:optimization} we report the interaction strength of different design based on the values of the device studied in this work.  In particular, we calculate the expected bound state interaction for the same parameters of the current sample, but increasing the free sites between the coupling points of the qubits. In the new configuration the qubits are coupled to cavity 9 and 13, instead of 10 and 12. The same calculation is repeated, but reducing the qubit-cavity coupling to 50\,MHz. With the latter configuration, we expect an on-off ratio of 1000 within 1\,GHz range.

\subsection{Electromagnetic cross-talk}
\label{app:crosstalk}

\begin{figure*}%
\includegraphics[width=\linewidth]{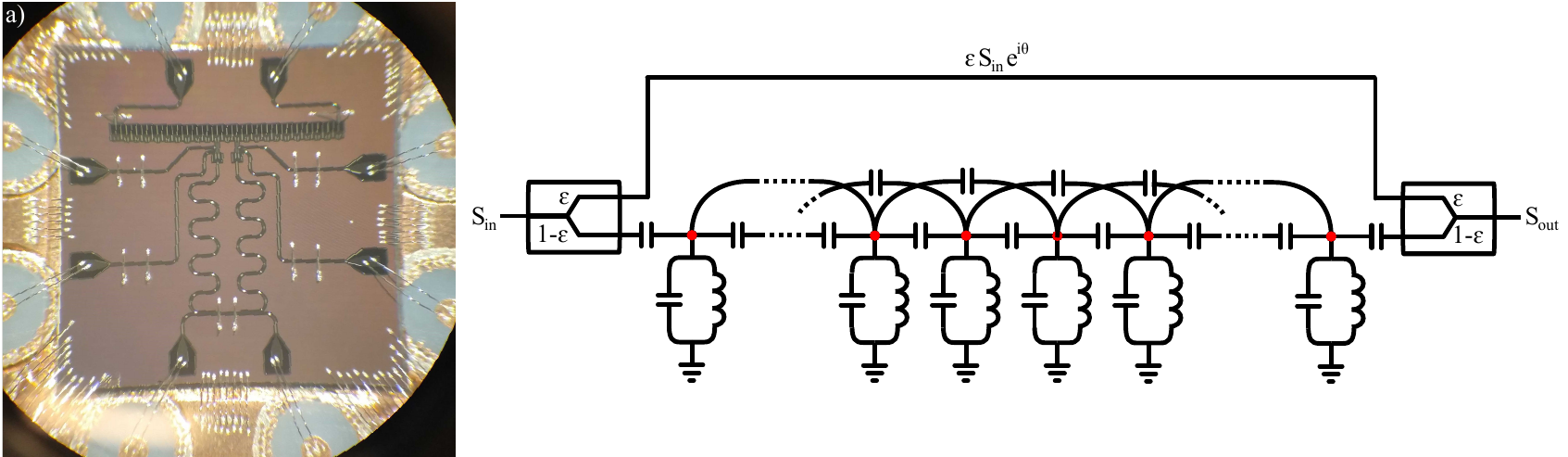}
\includegraphics[width=\linewidth]{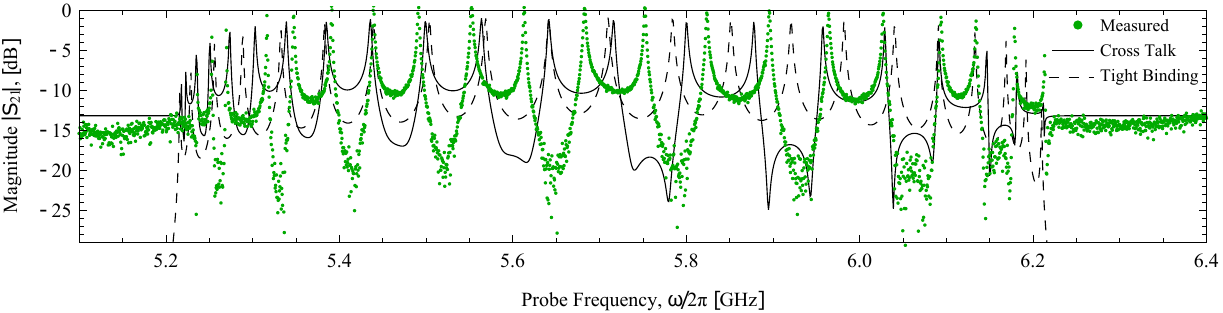}
\caption{Transmission cross-talk. (a) Micrograph of the sample bonded in the copper sample box. The 1.5\,mm aluminium bond wires connecting the inner conductor of a coaxial cable to the bond pad on the chips have a direct electromagnetic cross talk. (b) The model of the cross talk consists in two power splitter / combiner: a small portion of the signal bypass the waveguide and is recollected at the output, constructively or destructively interfering with the signal through the waveguide, and producing a non-zero transmission outside the band. (c) The measured transmission through the sample (green dots) is not well reproduced considering only the ideal case (input-output theory at low power, red dashed line). The sum on the signal through the array and the cross talk (black solid line) explain the non zero transmission and the "pairing" effect. 
\label{fig:crosstalk}}
\end{figure*}
Four features of the transmission spectrum across the coupled-cavity array cannot be quantitatively explained by the ideal case. The transmitted signal reported in \fig \ref{fig:crosstalk}(c) (compare it with \fig \ref{fig:naked_array} plotted with a linear scale), highlights the non vanishing transmission outside the band, the "pairing" effect of modes, the transmission minima within the band and finally the asymmetric mode distribution. These non-idealities are due to the non-negligible cross-talk between the input and output ports of the sample box and the next nearest neighbor interaction discussed in the section \ref{app:parassitic}.

\fig \ref{fig:concept}(a) reports a micrograph of the sample bonded to its sample holder. The bond wires connecting the sample box ports to the signal launcher on chip are approximately 1.5\,mm long. We model their electromagnetic cross-talk as two power dividers, the first of which separates the incoming signal $S_{\rm in}$ into a part that reaches the second power divider after an electrical delay $e^{i\theta}$ and a part through the coupled-cavity array, $\epsilon S_{\rm in}$. In our model the two signals are then recombined by the second power divider. 

The solid line in \fig \ref{fig:crosstalk}c shows the prediction for our model with the fitting parameters $\epsilon =0.22$ and $\theta= 0.34\pi$. In comparison with a pure tight binding model represented by the dashed black line, we can reproduce the main features of the transmission spectrum. Unfortunately, the mode distribution is still not completely described by the model. We attribute this behavior to fabrication imperfection, mainly on the smallest resonator feature represented by the junctions in the resonators.
The design of the JJ resonator intrinsically mitigates the possible variance by a factor of $\sqrt{10}$. Moreover, although the individual modes are affected by this imperfection, collective behavior of the coupled-cavity array as a waveguide is still close to the ideal because of the large resonator-resonator coupling $J$.

We would like to stress that the cross-talk affects only the measured transmitted field and not the intrinsic mode structure.


\subsection{Magnetic cross-talk}
The flux lines Z-\Qone and Z-\Qtwo have a linear magnetic cross-talk that has been calibrated and compensated. The flux reported here is therefore the net flux in each SQUID loop, and not the one produced by the aforementioned flux lines. 

The relation between the flux in each SQUID, $\Phi_{q1}$ and $\Phi_{q2}$, and the room temperature voltages applied to the coaxial cable, $V_1$ and $V_2$ respectively, is expressed by:
\begin{equation}
\left( \begin{array}{c}
\Phi_1  \\
\Phi_2  \end{array} \right) = 
\left( \begin{array}{cc}
L_{11} & L_{12}  \\
L_{21} & L_{22}  \end{array} \right)
\left( \begin{array}{c}
V_1  \\
V_2  \end{array} \right)+
\left( \begin{array}{c}
\Phi_1^{off}  \\
\Phi_2^{off}  \end{array} \right)
\end{equation}
where we introduced the inductance matrix for the flux lines $\mathbf{L}_{\rm fl}$. From the measurement of the periodicity of the readout resonator of each qubit at different flux we can extract the coefficients of the inductance matrix and the flux offest in each SQUID loop. 


In order to decouple the two flux lines, we can redefine the voltages in eache of them as:
\begin{equation}
    \left( \begin{array}{c}
V_1  \\
V_2  \end{array} \right) = 
\left( \begin{array}{cc}
L_{11} & L_{12}  \\
L_{21} & L_{22}  \end{array} \right)^{-1}
\left( \begin{array}{c}
\Phi_1 -\Phi_1^{\rm off} \\
\Phi_2 - \Phi_2^{\rm off}
\end{array} \right)
\end{equation}

For our sample we measure an offset $\Phi_1^{\rm off}/\Phi_0=0.091$ and 
$\Phi_2^{\rm off}/\Phi_0=0.084$, while the relative magnetic cross-talk for \Qone is $L_{12}/L_{11}=0.041$ and for \Qtwo is $L_{21}/L_{22}=0.063$.

\bibliography{biblio}

\end{document}